\documentclass[acmtog]{acmart}

\AtBeginDocument{%
  \providecommand\BibTeX{{%
    \normalfont B\kern-0.5em{\scshape i\kern-0.25em b}\kern-0.8em\TeX}}}

\copyrightyear{2025}
\acmYear{2025}
\setcopyright{acmlicensed}\acmConference[SIGGRAPH Conference
Papers '25]{Special Interest Group on Computer Graphics and
Interactive Techniques Conference Conference Proceedings}{August
10--14, 2025}{Vancouver, BC, Canada}
\acmBooktitle{Special Interest Group on Computer Graphics and
Interactive Techniques Conference Conference Proceedings (SIGGRAPH 
Conference Papers '25), August 10--14, 2025, Vancouver, BC, Canada}
\acmDOI{10.1145/3721238.3730638}
\acmISBN{979-8-4007-1540-2/2025/08}

\acmDOI{10.1145/3721238.3730638}

\usepackage{color}
\usepackage{xcolor}
\usepackage{amsthm, amsfonts, amssymb}
\usepackage{makecell}
\usepackage{textcomp}
\usepackage{epstopdf}
\usepackage{multirow}
\usepackage{wrapfig}
\usepackage{subfig}
\usepackage[ruled]{algorithm2e} 
\usepackage{cleveref}
\usepackage{bm}
\usepackage{enumitem}

\SetAlFnt{\small}
\SetAlCapFnt{\small}
\SetAlCapNameFnt{\small}
\SetAlCapHSkip{0pt}

\definecolor{green}{rgb}{0, 0.5, 0}
\definecolor{orange}{rgb}{0.6, 0.3, 0.1}
\definecolor{red}{rgb}{1.0, 0.0, 0.0}
\definecolor{teal}{rgb}{0.0, 0.4, 0.4}
\definecolor{purple}{rgb}{0.65,0,0.65}
\definecolor{saffron}{rgb}{0.95,0.75,0.2}
\definecolor{turquoise}{rgb}{0.0,0.5,0.5}
\definecolor{brown}{rgb}{0.5, 0.16, 0.16}
\definecolor{brickred}{rgb}{.6, .2 .1}
\definecolor{coral}{rgb}{1,0.45,0.33}
\definecolor{newcolor}{rgb}{.8,.349,.1}

\begin{document}
\title{CLR-Wire: Towards Continuous Latent Representations for 3D Curve Wireframe Generation}

\author{Xueqi Ma}
\email{qixuemaa@gmail.com}
\affiliation{%
	\department{College of Computer Science \& Software Engineering}
	\institution{Shenzhen University}
	\country{China}	
}

\author{Yilin Liu}
\email{whatsevenlyl@gmail.com}
\affiliation{%
	\institution{Shenzhen University}
	\country{China}	
}

\author{Tianlong Gao}
\email{gao.alone123@gmail.com}
\affiliation{%
	\institution{Shenzhen University}
	\country{China}	
}

\author{Qirui Huang}
\email{qrhuang2021@gmail.com}
\affiliation{%
	\institution{Shenzhen University}
	\country{China}	
}

\author{Hui Huang}
\email{hhzhiyan@gmail.com}
\authornote{Corresponding author: Hui Huang (hhzhiyan@gmail.com)}
\affiliation{%
	\department{College of Computer Science \& Software Engineering}
	\institution{Shenzhen University}
	\country{China}	
}

\renewcommand\shortauthors{X. Ma, Y. Liu, T. Gao, Q. Huang, and H. Huang}

\begin{abstract}
We introduce CLR-Wire, a novel framework for 3D curve-based wireframe generation that integrates geometry and topology into a unified \textit{Continuous Latent Representation}. Unlike conventional methods that decouple vertices, edges, and faces, CLR-Wire encodes curves as Neural Parametric Curves along with their topological connectivity into a continuous and fixed-length latent space using an attention-driven variational autoencoder (VAE). This unified approach facilitates joint learning and generation of both geometry and topology.
To generate wireframes, we employ a flow matching model to progressively map Gaussian noise to these latents, which are subsequently decoded into complete 3D wireframes. 
Our method provides fine-grained modeling of complex shapes and irregular topologies, and supports both unconditional generation and generation conditioned on point cloud or image inputs. 
Experimental results demonstrate that, compared with state-of-the-art generative approaches, our method achieves substantial improvements in accuracy, novelty, and diversity, offering an efficient and comprehensive solution for CAD design, geometric reconstruction, and 3D content creation.
\end{abstract}

%
%

\ccsdesc[500]{Computing methodologies~Shape modeling}

%
%

\keywords{3D Curve, Wireframe, 3D Generation, Flow Matching}

\begin{teaserfigure}
    \centering
    \includegraphics[width=0.99\linewidth]{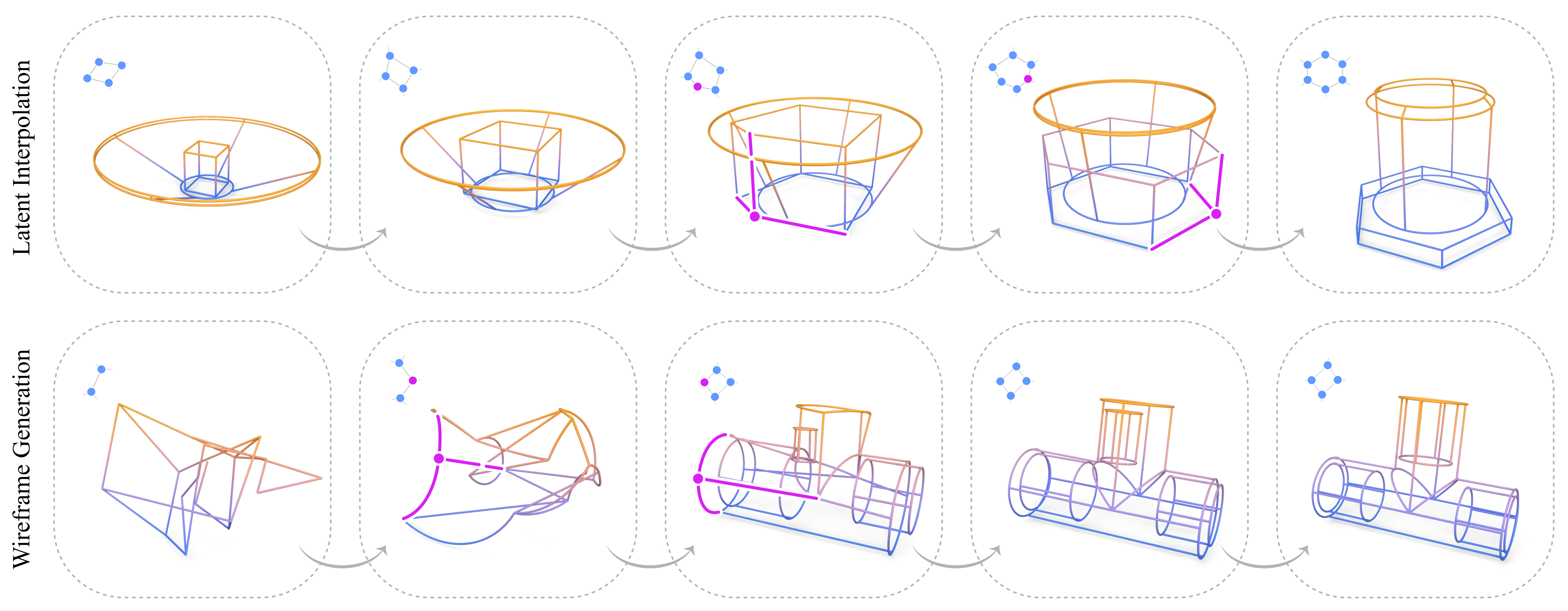}
    \caption{
        Our approach encodes discrete topological structures and continuous geometric information into a unified Continuous Latent Representation, enabling 3D wireframe interpolation and generation.
        \textit{Top.} Given two wireframes with different topology, our Continuous Latent Representation enables a smooth interpolation on both their geometry and topology.
        \textit{Bottom.} Starting from an initial noise, our method gradually generate a complex wireframe using a flow matching model.
        The small graphs in the upper-left corner of each sample illustrate the changes in topology during the process.
    }
    \label{fig:teaser}
\end{teaserfigure}

\maketitle

\section{Introduction}
3D curve wireframes, also known as curve networks, are simple yet powerful representations of 3D shapes, which provide concise abstractions of objects' fundamental structures.
They capture both sharp and non-sharp edge (descriptive curve~\cite{flowrep}) features of 3D shapes, serving as strong guidance for tasks such as shape abstraction~\cite{abstraction2009Mehra} and surface reconstruction~\cite{pan2015curve}.
Additionally, the low complexity of wireframes makes them suitable for interactive editing and communication among designers~\cite{iwires}.
However, as a structured representation, a typical 3D curve wireframe consists of both continuous wires and their discrete topological connectivity, making it challenging to be represented by a continuous network.

Early research on wireframe generation primarily focused on extracting wireframes from high-quality point clouds~\cite{PC2WF21,nerve}. 
These methods typically rely on learning-based or optimization techniques to detect corner points or edge points in 3D models. 
However, the topology reconstruction is usually missing in these methods, which typically results in a complex and parameter-sensitive post-processing step to obtain the final wireframe.
Recent generative models like SolidGen~\cite{solidgen23} and BrepGen~\cite{brepgen} solve this problem by generating B-rep CAD models using multiple models to represent vertices, edges, and faces separately.
Additionally, 3DWire~\cite{3dwire} directly models straight-line segments in an autoregressive manner to generate 3D wireframes.
However, they are primarily designed for regular 3D models or simple geometric shapes, and exhibit limited performance when dealing with irregular topologies and complex shapes.
Furthermore, 3D curve-based wireframes involve a complex combination of continuous and discrete structures, where geometry and topology are interdependent.
Existing methods that treat geometry and topology separately may introduce errors in distribution learning, thereby affecting overall performance.

We propose a novel 3D curved wireframe generation framework that jointly models the geometry and topology of 3D wireframes and maps them into a unified continuous latent space.
Curves are first normalized and represented as smooth Neural Parametric Curves via a curve encoder.
Then we use several cross attention modules to encode both the Neural Parametric Curves and their topological connectivities into a unified and fixed-length latent space using a variational autoencoder(VAE)~\cite{vae}.
The latent space is then used to generate 3D wireframes both unconditionally and conditionally using a Flow Matching approach~\cite{flowmatching}.
Experimental results demonstrate that our proposed method achieves superior accuracy and diversity compared to state-of-the-art generative models and shows remarkable adaptability to complex shapes and topologies.

In summary, we make the following contributions: 

• Introducing a fixed-length and Continuous Latent Representation for jointly encoding the geometry and topology of 3D wireframes while allowing a smooth interpolation in the latent space.

• Proposing a 3D curve-based wireframe generation framework based on the smooth latent space which can effectively learn the distribution of complex 3D curve-based wireframes and significantly improve generation accuracy. 

• Enabling both unconditional and condition-driven 3D curve wireframe generation using point cloud or image inputs.

\section{Related Work}\label{sec:rw}

\subsection{Wireframe Generation}

\paragraph{Straight-line Wireframe Generation}
Recent advances~\cite{pbwr2024,3dwire,xue2024neat} in 3D wireframe generation have primarily focused on straight-line structures~\cite{Sugihara1982}, employing both image-based~\cite{how3d,Pautrat2023} and point cloud-based~\cite{PC2WF21} approaches. 
Image-based methods utilize 2D keypoints and line segments~\cite{zhou2019iccv,Lin2020eccv,huang2018} to infer 3D wireframes, with improvements in optimization through techniques like global attraction fields~\cite{Xue2023tpami}. 
Researchers have also integrated special structures like 3D skeletons~\cite{wu2018} and Manhattan wireframes~\cite{zhou2019learning} to enhance wireframe reconstruction.
Similarly, point cloud-based approaches~\cite{PC2WF21,Jung2016,Hofer2017} detect corner points and edges to construct topological graphs or refine geometries~\cite{Tan2022CG}. 
As for specialized tasks, methods like Point2Roof~\cite{wichmann2019roofn3d} and LC2WF~\cite{luo2022bmvc} have been developed for architectural applications.
However, these methods are inherently constrained to linear structures and lack adaptability to curve-based wireframes, limiting their applicability for complex geometries.

\paragraph{Curve-based Wireframe Generation}
Curve-based wireframe generation tackles the challenges of reconstructing wireframes with continuous curves. 
FlowRep~\cite{flowrep} generates compact, descriptive 3D curve networks by aligning flowlines with principal curvature, focusing on perceptual and geometric reconstruction.
Engineering drawing-based approaches~\cite{Zhang2004Reconstruction,zhang2023automatic} generate 3D wireframes from 2D orthographic projections, using pattern matching and cycle detection to improve accuracy. 
FaceFormer~\cite{faceformer} reformulates face recognition in 2D wireframe projections as a curve sequence generation problem, establishing geometric constraints between curves and patches.
Additionally, MV2Cyl~\cite{mv2cyl} and NEF~\cite{NEF} specialize in extruded cylinder models and parameterized Bézier curves, respectively, but remain sensitive to occlusions and missing edge information.
Point cloud-based approaches\cite{complexgen22,WireframeNet} primarily focus on feature line extraction and parameterized curve fitting.
Since sharp features are essential for wireframe generation, methods use various mechanisms like distance field functions~\cite{Matveev21Wireframe,DEF}, multi-stage processes or end-to-end architectures~\cite{rfeps,SepicNet}, implicit edges~\cite{nerve} and region proposal networks~\cite{pie-net} to identify feature regions and reconstruct curves.
Despite these advancements, these methods often decouple geometry and topology, limiting their ability to model complex structures.

\subsection{3D Shape Generation and Interpolation}
Recent advancements in 3D generative models have explored various 3D representations, ranging from explicit meshes~\cite{meshdiffusion,PolyDiffuse,PolyDiff}, implicit functions~\cite{CLAY,3dshape2vecset,SDFusion,LAS-Diffusion}, and enhanced point clouds~\cite{DiffFacto,dit-3d,zheng2024pointdif,pointflow,dpm-3d}.
Prior works have explored 3D shape interpolation by learning continuous latent spaces, either through signed distance functions~\cite{deepsdf, jingyu2024neuralwavelet}, disentangled geometry-structure representations~\cite{DSG-Net}, or unified structural and geometric modeling via hierarchical VAEs~\cite{SDM-NET}.
Generative models for Computer Aided Design (CAD) like SolidGen~\cite{solidgen23} and BrepGen~\cite{brepgen} generate B-rep CAD models by separately representing vertices, edges, and faces.
Despite the advancements in representing 3D geometry and topology simultaneously, the topology between primitives has been modeled separately from the geometry, which may lead to shape inconsistency during the generation process.

In contrast, our method employs Flow Matching~\cite{flowmatching} to directly model the joint geometry and topology distribution within a unified, continuous latent space. 
This enables efficient and smooth generation and interpolation of 3D curve-based wireframes, overcoming the limitations of discrete diffusion processes and decoupled modeling strategies.

\begin{figure*}
    \centering
    \includegraphics[width=0.99\linewidth]{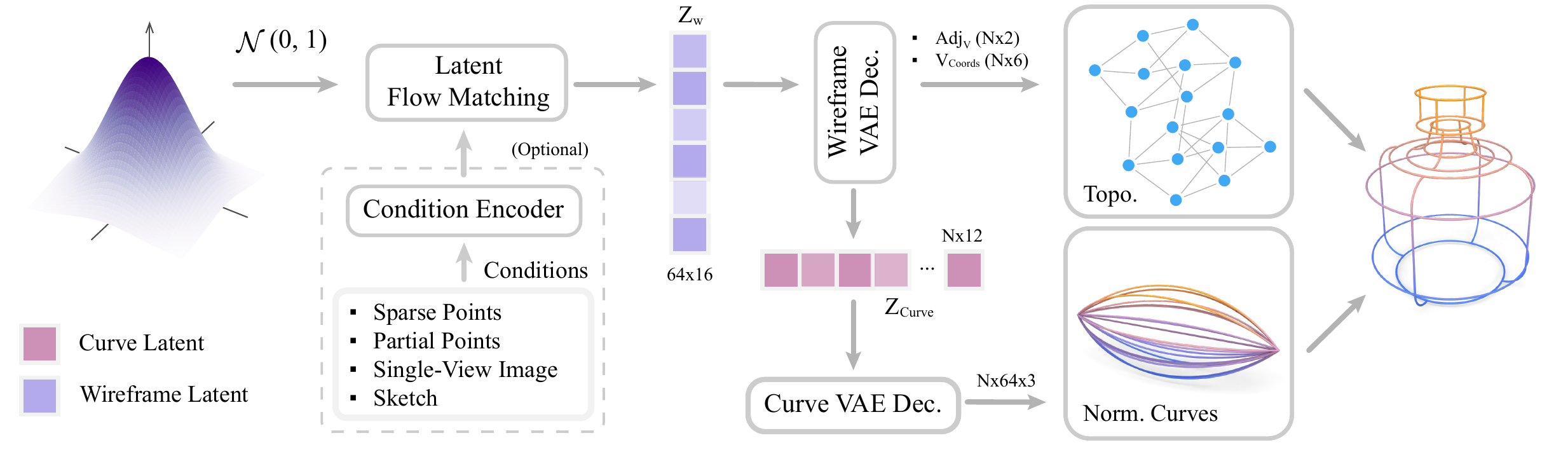}
    \caption{
    \textbf{Overview of wireframe generation.}
    Our method employs Latent Flow Matching based on the proposed latent wireframe representation $Z_W$, 
    which is decoded into a wireframe via the Wireframe Decoder.
    First, random noise is mapped to $Z_W$, and then decoded into adjacency $\text{Adj}_{V}$, 
    endpoint coordinates $V_{\text{Coords}}$, and curve latents $Z_{\text{Curve}}$.
    The final wireframe is produced using $\text{Adj}_{V}$, $V_{\text{Coords}}$, 
    and 3D curves decoded from $Z_{\text{Curve}}$.
    Furthermore, our method supports conditional inputs (e.g., sparse point clouds, partial point clouds, 
    single-view images, or sketches), offering controllable and flexible 3D generation.
    }
    \label{fig:overview}
\end{figure*}

\section{Methodology}\label{sec:method}

A 3D curve-based wireframe $\mathcal{W}(\{l_i\},A)$ consists of continuous geometric curves $\{l_i\}$ and their discrete topological connections $A$.
Unifying these two distinct data representations into a continuous latent space, capable of simultaneously modeling geometric and topological information, poses a significant challenge.
To address this issue, we propose a 3D curve wireframe generation framework (see Fig.~\ref{fig:overview}).
The framework consists of three components: 
1) Curve VAE to encode 
the various types of geometric curves $l$ (e.g., lines, circles, hyperbolas, Bézier curves, etc.) into a latent representation \(Z_{\text{Curve}}\); 
2) Wireframe VAE (see Fig.~\ref{fig:wireframe_vae}) to integrate the curve latents \(Z_{\text{Curve}}\) and their topological connections \(A\) into a global and fixed-length latent representation \(Z_W\); 
3) Latent Flow Matching to generate 3D wireframes, either unconditionally or conditionally.

\subsection{Curve VAE}
\label{sec:curve_vae}

\begin{figure}[t]
    \centering
    \includegraphics[width=0.99\linewidth]{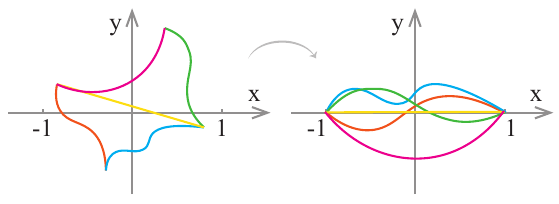}
    \caption{
        \textbf{2D Curve normalization examples}. Curves are normalized successively through translation, rotation, and scaling to align their start and end points, ensuring consistent spatial representation.
    }
    \label{fig:wire_norm}
\end{figure}

\begin{figure*}
    \centering
    \includegraphics[width=0.99\linewidth]{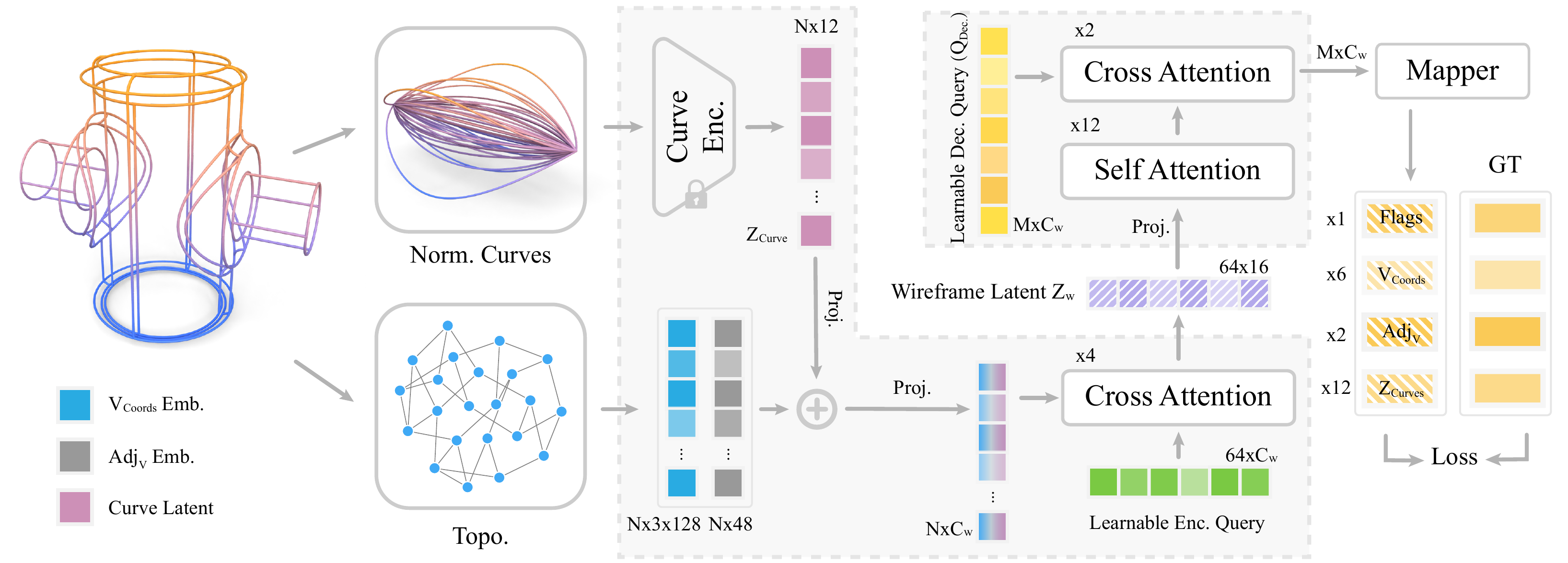}
    \caption{
    \textbf{Wireframe VAE pipeline}. 
        Given a 3D curve wireframe, we extract curve latents $Z_\text{Curve}$ from normalized curves via Curve Encoder, and the topology embeddings, which include adjacency $\text{Adj}_{V}$ embeddings and corresponding vertex coordinate $V_{\text{Coords}}$ embeddings.
        They are concatenated and transformed into a unified latent space $Z_W$ via Transformer blocks. In decoding phase, the $Z_W$ are decoded back to $\text{Adj}_{V}$, $V_{\text{Coords}}$, $Z_\text{Curve}$ and valid flags, which are used to compute losses against ground truths.
    }
    \label{fig:wireframe_vae}
\end{figure*}

\paragraph{Normalization.}  
Each curve $l$ is normalized to a standard format, where the start and end points are consistently set at [-1, 0, 0] and [1, 0, 0], respectively.
This process reduces redundancy and enables the network to handle curves of varying types and scales, thereby enhancing training stability.
For clarity, we employ 2D examples to explain the normalization process without loss of generality. As shown in Fig.~\ref{fig:wire_norm}, it consists of three steps:
    1) Translation: Shift the curve so that its first endpoint is at [-1, 0];  
    2) Rotation: Align the curve’s second endpoint with the position [1, 0];  
    3) Scaling: Adjust the curve's scale to ensure consistency across different instances.

\paragraph{Encoding.}

The normalized curve is encoded into a latent representation \( Z_{\text{Curve}} \).
First, we sample 256 points \( X \in \mathbb{R}^{256 \times 3} \) along the curve to ensure a well represent of complex curves.
Unlike BrepGen~\cite{brepgen}, which directly encodes the curve using convolutional and downsampling layers, we first reduce the spatial dimension from 256 to 64 using a cross-attention mechanism~\cite{3dshape2vecset}, where the key and value are \(X\), and the query is a subset \(X^* \in \mathbb{R}^{64 \times 3}\) sampled from \(X\).
This step yields \(\phi_\text{Curve} \in \mathbb{R}^{64 \times C}\):
\[
\phi_{\text{Curve}} = \text{CrossAttn}\bigl(\text{PosEmb}(X^*), \text{PosEmb}(X)\bigr), \\
\]
where $\text{PosEmb}$ is the positional encoding function and $C=64$.
Then, \(\phi_\text{Curve}\) is compressed into \(Z_{\text{Curve}} \in \mathbb{R}^{4 \times 3}\) 
through a stack of 1D convolutional layers.
This process provides a compact curve representation for the subsequent Wireframe VAE, while effectively preserving the curve’s geometric details.

\paragraph{Decoding.}

The latent curve representation \(Z_{\text{Curve}}\) is decoded into the geometric curve \(\hat{l}=\hat{l}(t)\).
First, a 1D upsampling convolutional network transforms \(Z_{\text{Curve}}\) into features \(\hat{\phi}_\text{Curve} \in \mathbb{R}^{64 \times C}\). 
Next, we introduce a query variable \(t\), which we embed via \(\mathrm{PosEmb}(t)\), a Fourier feature embedding, and fuse with \(\hat{\phi}_\text{Curve}\) through cross-attention.
The result for a given \(t\) is then obtained as follows:
\[
\hat{l}(t) = \mathrm{MLP}\Bigl(\mathrm{CrossAttn}\bigl(\mathrm{PosEmb}(t),\, \hat{\phi}_\text{Curve}\bigr)\Bigr),
\]
where \(\mathrm{MLP}\) is a multi-layer perceptron. 
In this manner, each query \(t\) is mapped to a unique 3D point \(\hat{l}(t)\) on the curve, yielding a continuous, parametric reconstruction from the compact representation. 
More details can be found in the supplementary material.

\subsection{Wireframe VAE}
\label{sec:wireframe_vae}

\paragraph{Encoding.}

We encode both the wireframe's geometric information and topological structure into a unified and fixed-length latent space \(Z_W\).
The geometric information of the wireframe is represented by the curve feature \(Z_{\text{Curve}}\), which is extracted from \(\{l_i\}\) through the Curve VAE.
The topology structure is represented by an adjacency list \(\text{Adj}_V \in \mathbb{R}^{N \times 2}\), and corresponding vertex coordinates \(V_{Coords} \in \mathbb{R}^{N \times 6}\), where \(N\) is the number of curves.

Each row \(\text{Adj}_{V_i}\) represents a connection, storing the 
indices of the start and end vertices for the \(i\)-th curve.
To ensure a consistent ordering of these connections, we sort the adjacency list via a breadth-first search (BFS) traversal, which helps maintain a similar adjacency distribution across different wireframes (see the supplementary material for details).
We concatenate the curve features \(Z_{\text{Curve}}\), the endpoint coordinates $V_{Coords}$, and \(\text{Adj}_V\), 
then use a perceiver-based feature aggregation module~\cite{perceiver} to map the combination into a fixed-length latent representation \(Z_W\):
\[
Z_W = \text{Perceiver}\Bigl(
    \text{Emb}_c\bigl(Z_{\text{Curve}}\bigr),\;
    \text{Emb}_v\bigl(V_{\text{Coords}}\bigr),\;
    \text{Emb}_a\bigl(\text{Adj}_{V}\bigr)
\Bigr),
\]
where \(\text{Emb}_c\), \(\text{Emb}_v\), and \(\text{Emb}_a\) are embeddings for the curve features $Z_{\text{Curve}}$, endpoint $V_{\text{Coords}}$, and the adjacency list $\text{Adj}_{V}$, respectively.

\paragraph{Decoding.}
We decode the fixed-length latent representation \(Z_W\) to reconstruct the wireframe \(\hat{\mathcal{W}}\), 
where the number of curves can vary.
Specifically, we employ $M$ learnable query vectors $Q_{Dec}$ to apply cross-attention on $Z_W$, extracting features for the individual curves. 
The maximum number of curves in the wireframe is pre-defined as $M$, and an additional validity flag is introduced to determine whether the decoded curves are valid. 
These features are subsequently mapped by a feature mapper to produce the adjacency list $\text{Adj}_V$, vertex coordinates $V_{\text{Coords}}$, and curve features $Z_{\text{Curve}}$, thereby fully reconstructing the wireframe \(\hat{\mathcal{W}}\).

\paragraph{Differential adjacency list.}
Moreover, we leverage the observation that the sorted adjacency list forms an increasing sequence. This property allows us to simplify the adjacency list further by constructing a differential adjacency list $\Delta \text{Adj}_V$:
\[
\begin{aligned}
\Delta \text{Adj}_V[i,0] &= \text{Adj}_V[i,0] - \text{Adj}_V[i-1,0], \\
\Delta \text{Adj}_V[i,1] &= \text{Adj}_V[i,1] - \text{Adj}_V[i,0].
\end{aligned}
\]
The $\Delta \text{Adj}_V$ provides a compact representation of the wireframe topology, reducing redundancy and improving computational efficiency and prediction accuracy.
Additional details are provided in the supplementary material.

\paragraph{Loss functions.}
During training, we optimize the Wireframe VAE using three loss functions: 1) MSE loss for curve latent representations $Z_{\text{Curve}}$ and vertex coordinates $V_{\text{Coords}}$; 2) cross-entropy loss for predicting the adjacency list $\text{Adj}_{V}$; 3) KL divergence loss to regularize the wireframe representation $ Z_W $ to ensure the distribution aligns well for subsequent generative modeling.

\subsection{Latent Flow Matching}
\label{sec:lfm}

Based on the wireframe latent representation $Z_W$, we use the flow matching model ~\cite{flowmatching} to progressively map Gaussian noise to the wireframe latent distribution, either unconditionally or conditionally.
We train a neural network that represents the velocity field $U_\theta$, which describes the instantaneous velocity of the wireframe latent representation:
\[
    \dot{X}_t = U_\theta(t, X_t, c),
\]
where $X_t$ is the wireframe latent representation at time $t$, $\theta$ represents the learnable parameters of the neural network, and $c$ denotes an optional conditioning variable. 
The Gaussian noise distribution $p_0$ is progressively transformed into the target latent distribution $p_1$ by evolving along the probabilistic path $p_t$. 
The flow matching model is optimized by minimizing the following loss function:
\[
L = \mathbb{E}_{X_0 \sim p_0, t \sim \mathcal{U}[0,1]} \left[ \| \dot{X}_t - U_\theta(t, X_t, c) \|^2 \right],
\]
where $X_0$ is the initial input sampled from the Gaussian noise distribution $p_0$. 
See supplementary material for more details.
After training, new samples from the target distribution are generated by solving the ordinary differential equation (ODE) defined by the velocity field. 
For condition generation, we utilize a frozen DINOv2~\cite{dinov2}\ and a learned PointNet++~\cite{pointnet++} as feature extractors for processing images and point clouds, respectively.

\section{Experimental Results}
\label{sec:results}

\subsection{Implementation}

\paragraph{Data filtering.}
We conduct experiments using the ABC dataset~\cite{ABC}. We filter out multi-solid samples, simple shapes such as cuboids and cylinders, and overly complex samples with more than $128$ curves. Following SolidGen~\cite{solidgen23}, we employ a hash-based deduplication method to remove duplicate samples. This process results in a final dataset of $130,473$ samples, which is split into $90\%$ for training, $5\%$ for testing, and $5\%$ for validation.

\paragraph{Training details.}
The models were trained with a learning rate of 1e-4, employing a linear warmup phase and cosine decay scheduler, and using AdamW as the optimizer. For CurveVAE, the KL divergence loss weight was set to 5e-6, while for WireframeVAE, it was set to 5e-5.
We use a learned PointNet++~\cite{pointnet++} and DINOv2~\cite{dinov2} to extract point cloud and image embeddings, respectively.
These embeddings are then projected to $d=768$ feature vector and fed into the DiT backbone using the AdaLN-Zero~\cite{dit} injection. 
During training, we used a linear path for the flow matching trajectory in Latent Flow Matching. At inference, we sampled wireframe latent variables with an ODE solver over 250 steps, then fed the result into the WireframeVAE and CurveVAE decoders to generate the final curve wireframe.
All experiments were conducted on 8 NVIDIA RTX 4090 GPUs.
CurveVAE took about one day to train, WireframeVAE required four days, and the Latent Flow Matching model was trained over five days.
More details can be found in the supplementary material.

\paragraph{Metrics.}
To assess the performance of our method, we employed multiple evaluation metrics. Geometric similarity between reference wireframes \(\mathcal{W}\) and generated wireframes \(\hat{\mathcal{W}}\) was quantified using Chamfer Distance (CD) and Earth Mover's Distance (EMD). 
We also use distribution-based metrics to evaluate the diversity and fidelity of the generated samples: 
1) Coverage (COV) to assess the diversity of the generated samples;
2) Maximum Mean Discrepancy (MMD) to evaluate the fidelity of the results and 
3) 1-Nearest Neighbor (1-NN) to quantify the similarity between the distributions of \(\mathcal{W}\) and \(\hat{\mathcal{W}}\). 

\subsection{Unconditional Generation}
\begin{figure*}
    \centering
    \includegraphics[width=0.99\linewidth]{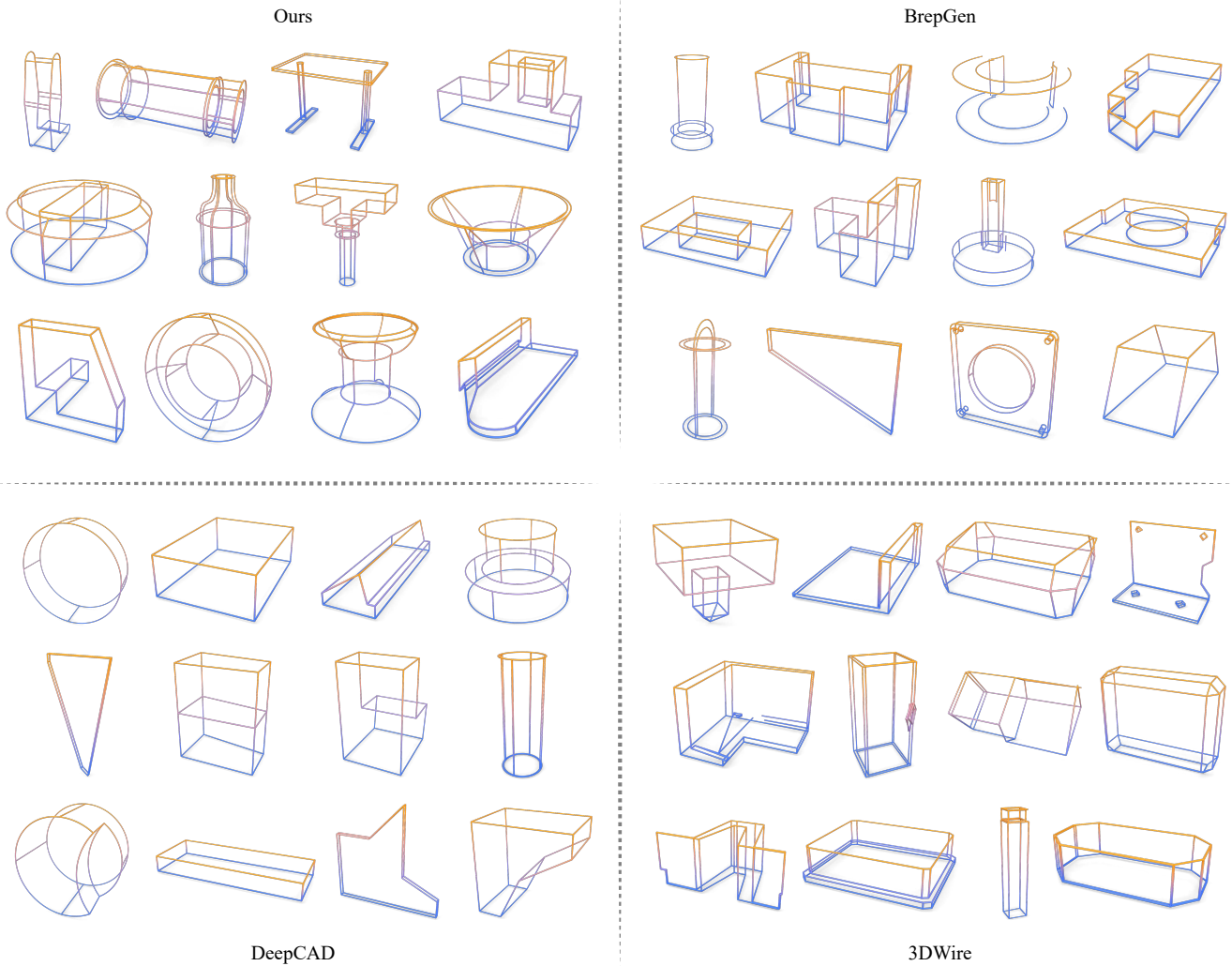}
    \caption{
    Qualitative evaluation of our method with state-of-the-art methods, BrepGen\cite{brepgen}, DeepCAD\cite{deepcad}, and 3DWire\cite{3dwire}.
    Our method generates wireframes with richer curve details, maintaining geometric integrity and topological correctness, especially for complex shapes.
    }
    \label{fig:uncond}
\end{figure*}

\paragraph{Baselines.}
We compare our method with the state-of-the-art wireframe generation method 3DWire~\cite{3dwire}.
Additionally, we include results from CAD generative models DeepCAD~\cite{deepcad} and BrepGen~\cite{brepgen} since a wireframe can also be extracted from their generated CAD models.
We use the pre-trained models of DeepCAD and BrepGen and retrain 3DWire using our curve wireframe dataset to ensure its compatibility with curve wireframe generation.

\paragraph{Qualitative evaluation}

Fig. \ref{fig:uncond} illustrates the unconditional wireframe results generated by different methods. 
3DWire cannot efficiently represent curves, resulting in wireframes with straight-line segments only.
While BrepGen and DeepCAD can generate CAD models with curved surfaces, they tend to produce simpler wireframes.
Their wireframe generation process is largely dependent on surface generation, which significantly limits the complexity of the wireframes.
In contrast, our approach directly and simultaneously encodes the topology and geometry into a unified continuous latent representation, enabling the generation of both complete wireframes and intricate freeform curves.
Additionally, Fig. \ref{fig:more_results} showcases additional samples of 3D curve wireframes generated by our method, showcasing its versatility and capability.

\paragraph{Quantitative evaluation}
\begin{table}[t!]
    
    \caption{
    Quantitative evaluation of unconditional generation on the ABC dataset~\cite{ABC}. Our method outperforms others on COV, MMD, and 1-NN.
    Note that CD and EMD values are multiplied by \(10^2\).
    }
    \label{tab:quantitative}
    \centering
    \setlength{\tabcolsep}{5pt}
    \resizebox{\linewidth}{!}
    {
    \begin{tabular}{@{}lcccccc@{}}
        \toprule
        \multicolumn{1}{c}{\multirow{2}{*}{Model}} & \multicolumn{2}{c}{COV (\%, \textuparrow)} & \multicolumn{2}{c}{MMD (\textdownarrow)} & \multicolumn{2}{c}{1-NN (\%)} \\
        \cmidrule(lr){2-3} \cmidrule(lr){4-5} \cmidrule(lr){6-7}
             & CD & EMD & CD & EMD & CD & EMD \\  
        \midrule
        DeepCAD \cite{deepcad}      & 30.59 & 31.93 & 3.94 & 10.72  & 75.34  & 79.49 \\
        BrepGen \cite{brepgen}  & 35.89 & 36.32 & 3.71 & 9.51  & 70.99  & 74.73 \\
        3DWire \cite{3dwire}  & 45.69 & 48.31 & 3.41 & 8.83  & 64.49  & 62.15 \\
        Ours                   & \textbf{48.30} & \textbf{50.34} & \textbf{3.07} & \textbf{8.36}  & \textbf{54.10}  & \textbf{54.98} \\
        \bottomrule
        
    \end{tabular}
    }
\end{table}

To quantitatively evaluate the performance of our method, we use COV, MMD, and 1-NN to compare the generated wireframes with the test set samples.
We generated 10k wireframe samples and performed 10 independent random samplings, selecting 2k samples in each iteration.
As shown in Table~\ref{tab:quantitative}, our method outperforms all baseline methods across all metrics based on CD and EMD.
The higher COV indicates greater diversity in generated wireframes, while the lower MMD reflects stronger geometric consistency. Moreover, the 1-NN metric shows that our method’s distribution closely matches the test set, underscoring its robust modeling capabilities.

\paragraph{Shape novelty analysis}

\begin{figure}[t]
    \centering
    \includegraphics[width=0.99\linewidth]{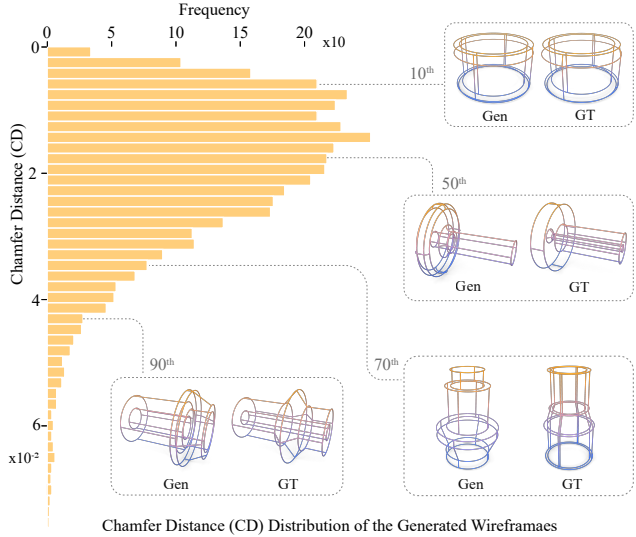}
    \caption{
        Novelty analysis.
        We present the Chamfer Distance (CD) distribution for 4k randomly generated wireframes compared to their most similar samples in the training set. Visualizations at various CD percentile ranges highlight both close resemblance (low CD) and novelty (high CD), showing that our method produces realistic and diverse shapes.
    }
    \label{fig:novelty}
\end{figure}

For each generated wireframe, we also retrieved the most similar sample in the training set by computing the minimum Chamfer Distance (CD) to evaluate the novelty of the generated samples.
Fig. \ref{fig:novelty} shows the distribution of the minimum CD values for 4k generated wireframes, along with the most similar training set samples for each generated wireframe.
More results can be found in the supplementary materials.
Additionally, we analyzed the topology of $1,000$ generated samples by checking if they have an isomorphic graph to those in the training set. The results show that $7.3\%$ of them have new topological structures.

\subsection{Conditional Generation}

\begin{figure}[t]
    \centering
    \includegraphics[width=0.99\linewidth]{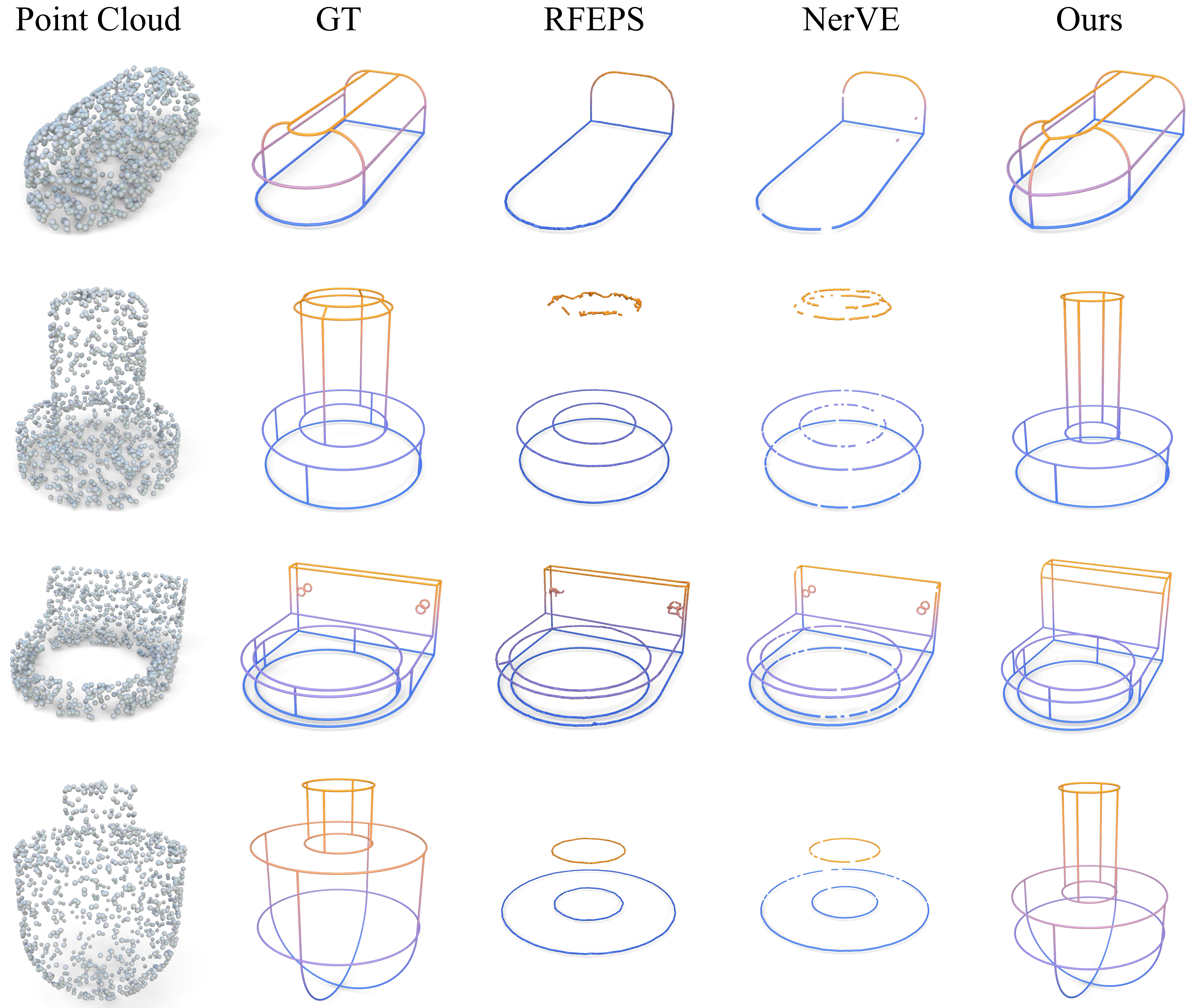}
    \caption{
    Comparison with RFEPS~\cite{rfeps} and NerVE~\cite{nerve} under the sparse point cloud condition. 
    Our method achieves competitive results even with fewer points.
    }
    \label{fig:cond_sparse}
\end{figure}

\begin{table}[t!]
    
    \caption{
    Quantitative evaluation of point cloud conditional generation. Our method outperforms others on CD, EMD and F-score.
    Note that CD is multiplied by \(10^3\) and EMD by \(10^2\).
    }
    \label{tab:pc_cond}
    \centering
    \setlength{\tabcolsep}{5pt}
    {
    \begin{tabular}{@{}lcccccc@{}}
        \toprule
        Method & CD(\textdownarrow) & EMD(\textdownarrow) & F-score(\textuparrow)\\ 
        \midrule        
        RFEPS~\cite{rfeps}     & 13.79 & 8.56 & 0.894 \\
        NerVE~\cite{nerve}     & 12.73 & 10.39 & 0.905 \\
        Ours                   & \textbf{8.26} & \textbf{2.58} & \textbf{0.910} \\
        \bottomrule
        
    \end{tabular}
    }
\end{table}

\begin{figure}[t]
    \centering
    \includegraphics[width=0.99\linewidth]{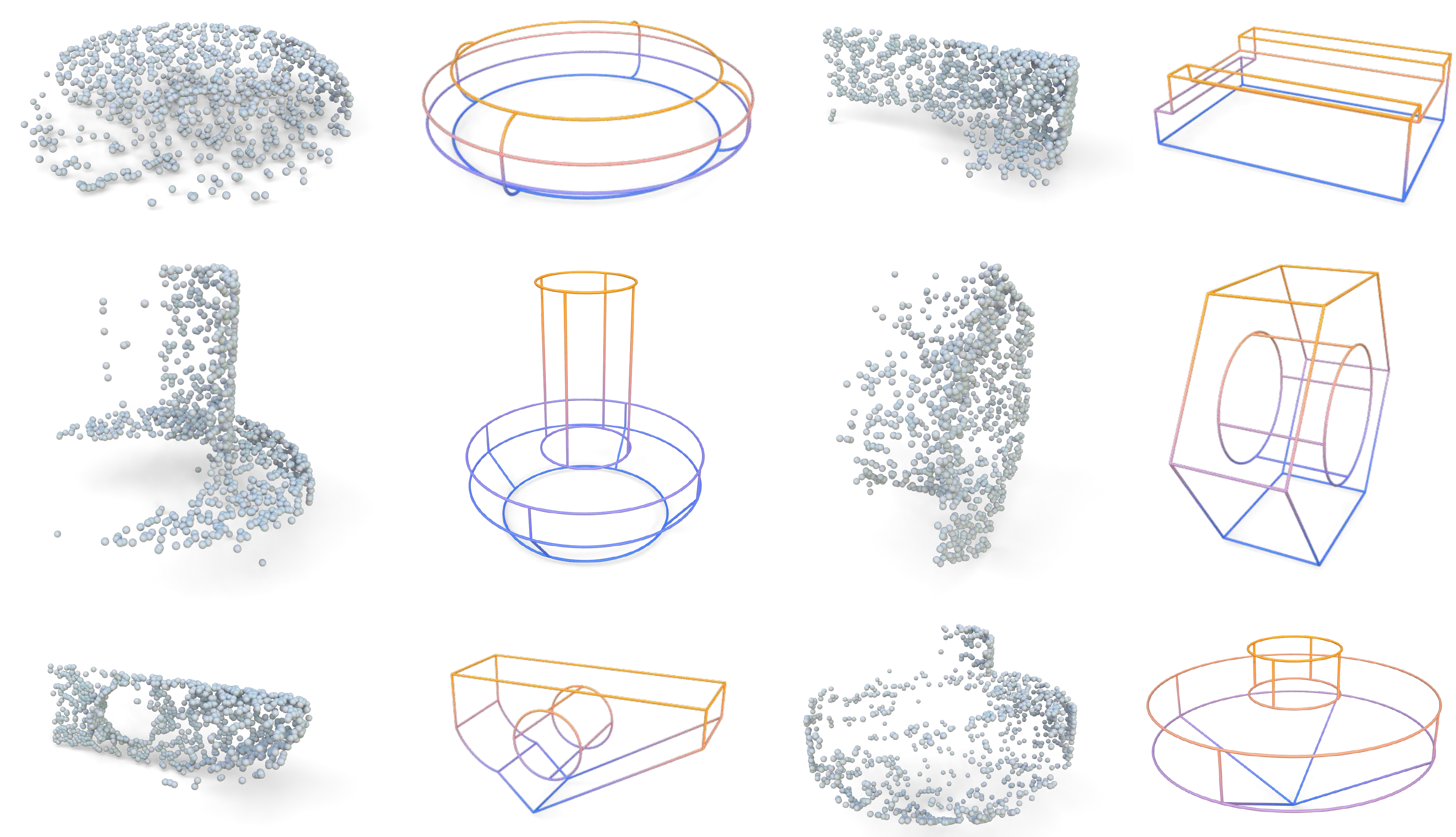}
    \caption{
    Visualization examples under the partial point cloud condition. 
    By leveraging the proposed latent wireframe representation, which integrates geometric and topological priors, our method generates geometrically consistent wireframes.
    }
    \label{fig:cond_partial}
\end{figure}

We evaluated generation capabilities of our method under various input conditions, focusing on three main types: point clouds, images, and sketches. 
The conditional generation model shares the same architecture, hyperparameters, and training strategy as the unconditional model.
Specifically, for point cloud inputs, we evaluated our method’s performance in low-quality scenarios (e.g., sparse and partial point clouds) to assess its robustness and flexibility.

We compare our point-conditioned wireframe generation with the state-of-the-art wireframe reconstruction methods, RFEPS~\cite{rfeps} and NerVE~\cite{nerve}.
Both methods require dense point clouds with 20K points for wireframe generation, while our method achieves competitive results using only sparse point clouds with 1K points as input conditions.
Quantitative results are presented in Table \ref{tab:pc_cond}, where our method demonstrates superior performance compared to the baseline methods in terms of Chamfer Distance (CD), Earth Mover’s Distance (EMD) and F-score (F1). Additionally, visual comparisons are provided in Fig. \ref{fig:cond_sparse} to illustrate the effectiveness of our approach.
Existing methods often rely on detecting sharp edges to reconstruct wireframe structures. However, for objects with smoother surfaces, these methods struggle to extract sufficient geometric features, leading to incomplete or inaccurate reconstructions. 
Furthermore, these methods rely on local geometric features and require dense point clouds for effective reconstruction. Sparse or incomplete inputs often lead to missing details and compromised structural quality.

In contrast, our method leverages a pre-trained latent space that efficiently captures global features while incorporating geometric and topological priors. This enables our approach to ensure the completeness of reconstruction and topological consistency of the generated wireframes. 
The results show that the wireframes generated by our method exhibit significantly better completeness of reconstruction than those produced by baseline methods.
Finally, as shown in Fig. \ref{fig:cond_partial}, our method proves robust even under partial point cloud conditions. It can generate complete and geometrically consistent wireframes, demonstrating its ability to handle sparse and incomplete inputs while maintaining high-quality results.

\begin{figure*}[t]
    \centering
    \includegraphics[width=0.96\linewidth]{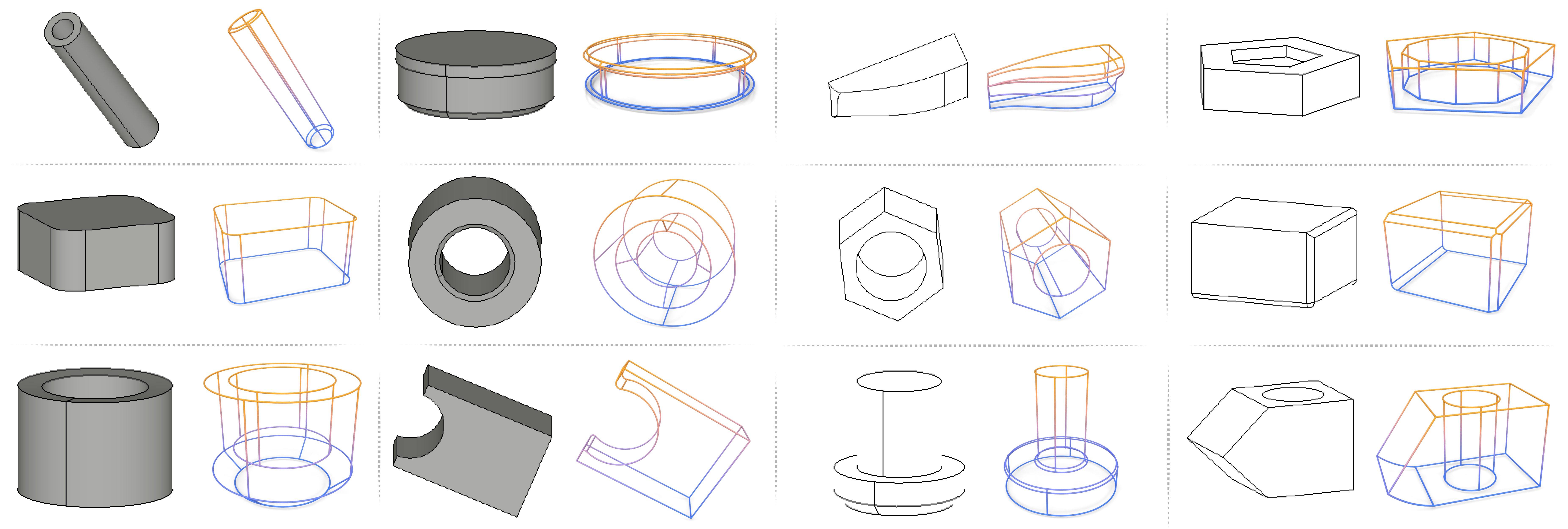}
    \caption{
    Visualization examples under single-view image and sketch conditions.
    }
    \label{fig:cond_image_sketch}
\end{figure*}

Furthermore, we tested images and sketches as conditional inputs for wireframe generation. The results are shown in Fig. \ref{fig:cond_image_sketch}, illustrating that our method is capable of producing complete and novel 3D curve wireframes under single-view image or sketch conditions. 
More discussions and limitations can be found in the supplementary.

\subsection{Latent Space Interpolation}

\begin{figure*}
    \centering
    \includegraphics[width=0.95\linewidth]{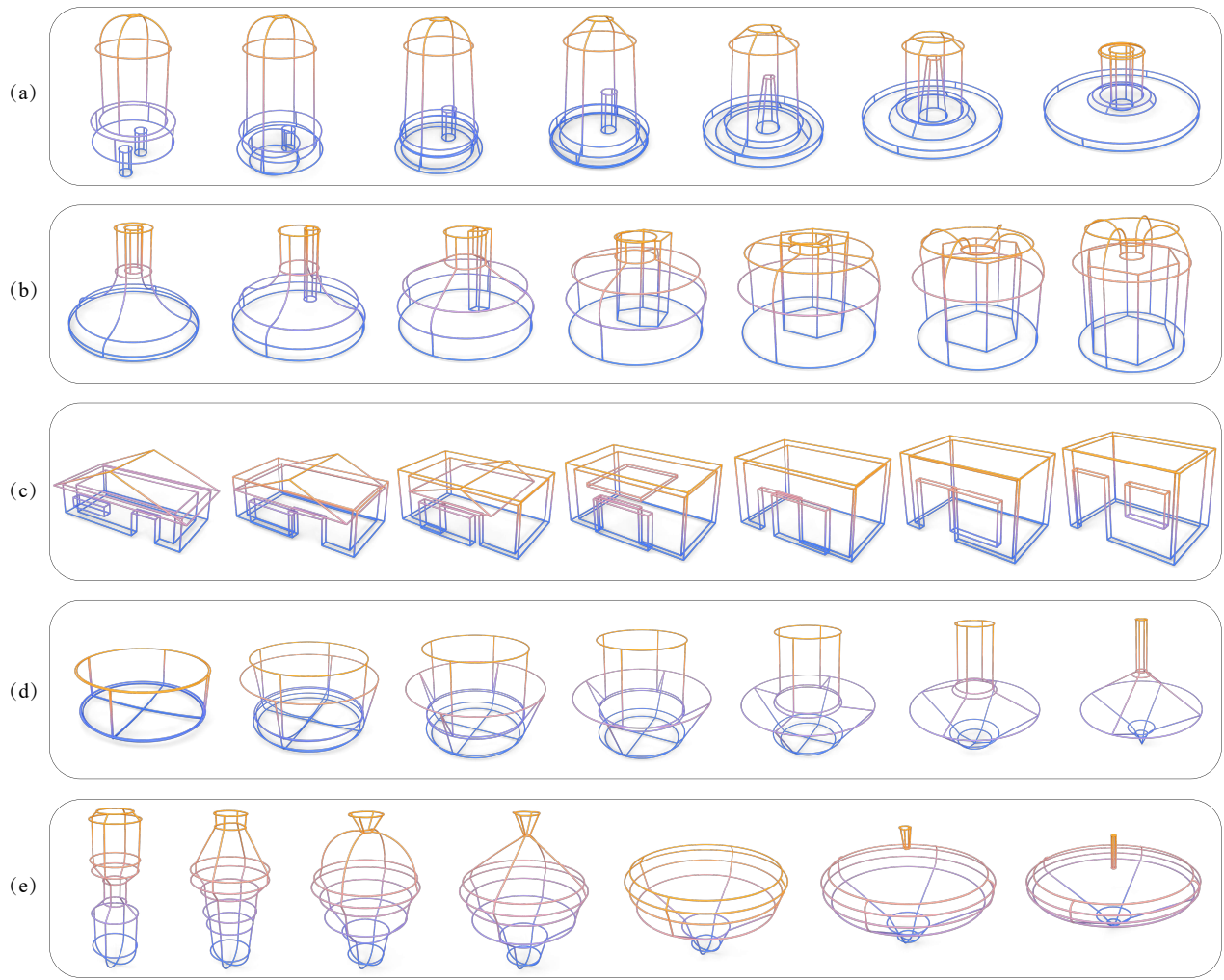}
    \caption{
    Visualization examples for latent interpolation.
    (b)-(e) show continuous transitions between shapes with different  topologies.
    Specifically, (e) illustrates a shape interpolation from a bottle-like structure to an open bowl-like form, showcasing the model’s ability to capture geometric and topological changes.}
    \label{fig:latent_interpolation}
\end{figure*}

\begin{figure*}[t]
    \centering
    \includegraphics[width=0.99\linewidth]{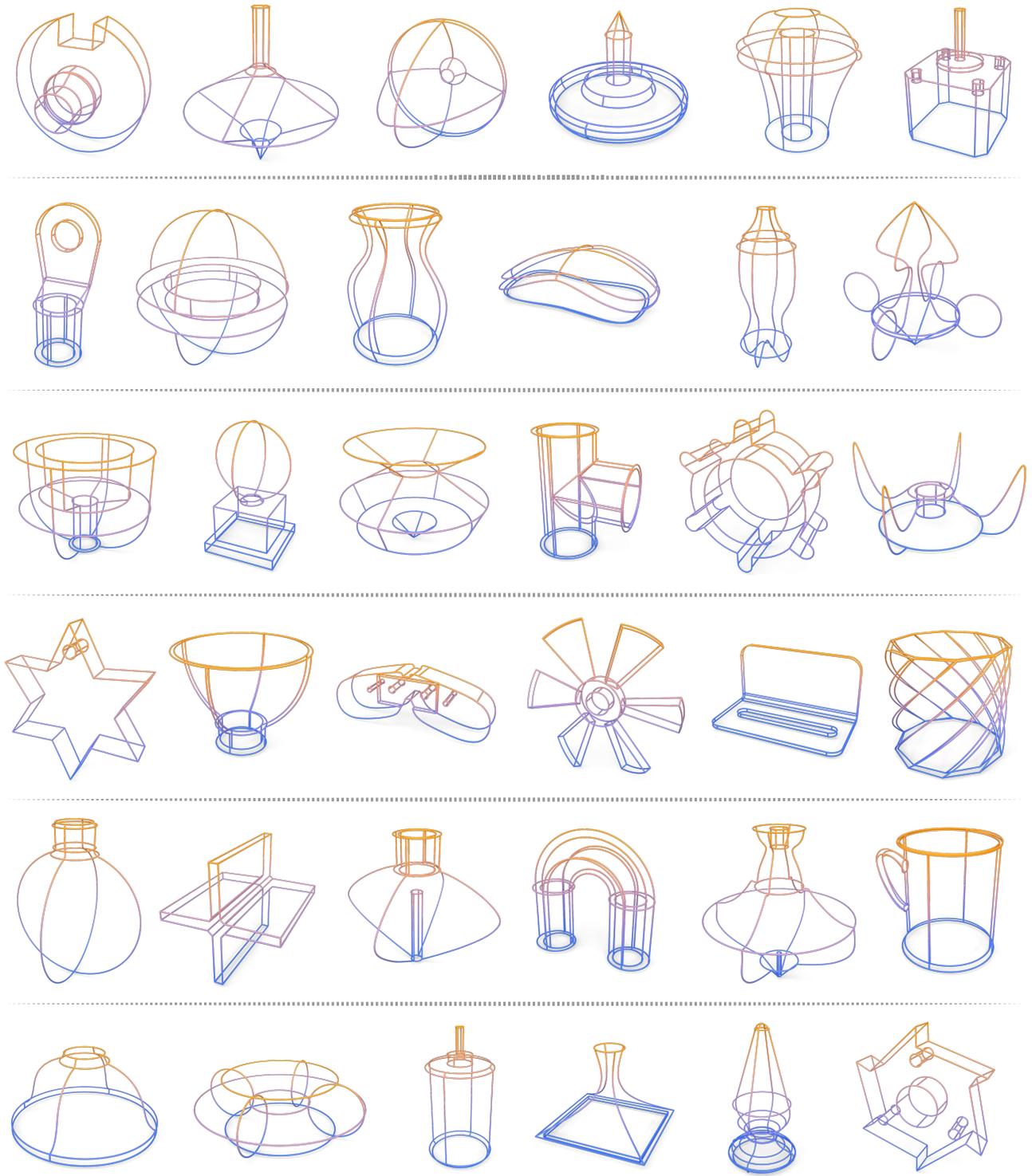}
    \caption{
    Visualization examples of unconditional wireframe generation.
    Our method is capable of generating high-quality wireframes with complex topological and geometric structures.
    }
    \label{fig:more_results}
\end{figure*}

By leveraging our continuous and fixed-length latent representation, we achieve smooth interpolation between two wireframes, as shown in Fig.~\ref{fig:latent_interpolation}.
Specifically, we extract the latent vectors corresponding to two distinct wireframes and employ spherical linear interpolation (slerp) to get a sequence of intermediate latent vectors. These interpolated vectors are subsequently decoded into 3D curve wireframes, resulting in a seamless shape transition between the given wireframes.
The interpolation not only preserves the fidelity of geometric details across intermediate shapes but also effectively captures topological variations between the initial and final wireframes (as shown in Fig.~\ref{fig:latent_interpolation}b–e), indicating its potential for future tasks like intuitive wireframe editing and manipulation.

\section{Conclusion and Future Work}
\label{sec:conclusion}

In summary, we propose a novel 3D curve wireframe model that significantly enhances the generation of complex curve wireframes. 
To the best of our knowledge, this is the first model that represents irregular curve wireframes within a continuous latent space.
Geometric details and topological connections are well blended in the latent space, ensuring consistency and integrity throughout the generation process.
The smooth interpolation in the latent space, superior generation diversity, and accuracy of condition generation demonstrate the expressiveness of the proposed representation.

Although our proposed latent space demonstrates smooth interpolation capabilities, further research is needed for controllable wireframe generation and editing. In the future, we aim to align the latent space more closely with text descriptions to achieve a more high-level semantic control of the generation process.

\begin{acks}
We thank all the anonymous reviewers for their insightful comments. 
Thanks also go to Xingguang Yan for helpful discussions. 
This work was supported in parts by National Key R\&D Program of China (2024YFB3908500, 2024YFB3908502), NSFC (U21B2023), ICFCRT (W2441020), Guangdong Basic and Applied Basic Research Foundation (2023B1515120026), DEGP Innovation Team (2022KCXTD025), Shenzhen Science and Technology Program (KJZD20240903100022028, KQTD20210811090044003, RCJC20200714114435012), and Scientific Development Funds from Shenzhen University.
\end{acks}

\bibliographystyle{ACM-Reference-Format}
\bibliography{clr-wire}


\clearpage

\appendix

This supplementary material includes technical explanations of our methods, extended visualizations of results, comparisons with other approaches, further analysis of shape novelty, quantitative evaluation metrics, and ablation studies. Finally, we discuss the limitations of our approach.

\section{Method Details}
\subsection{Curve VAE}

\paragraph{Normalization.}

To ensure a consistent representation of the curves, we first perform dense uniform sampling on each curve to obtain 256 3D points. This process transforms each curve into a polyline represented by a set of points. Such a normalized representation allows subsequent geometric transformations and analyses to handle the curves as structured point sets effectively.

To normailze the input curve, we employ the following steps:

1) Translation: Shift the starting point  $p_1$  of the polyline to $(0, 0, 0)$ to eliminate the influence of the initial position. The new set of curve points is represented as:
\[
\mathbf{P}' = \mathbf{P} - p_1.
\]

2) Rotation: Align the end vector of the curve $\mathbf{p}_n^{\prime} = p_n - p_1$ with the target direction vector $(1, 0, 0)$. Using Rodrigues’ rotation, we construct the rotation matrix $\mathbf{R}$:
\[
\mathbf{R} = \mathbf{I} + \sin\theta \mathbf{K} + (1 - \cos\theta)\mathbf{K}^2,
\]
where $\mathbf{K}$ is the skew-symmetric matrix derived from the rotation axis vector $\mathbf{a}$. The rotation axis $\mathbf{a}$ is calculated as:
\[ \mathbf{a} = \frac{\mathbf{p}_n' \times (1, 0, 0)}{\|\mathbf{p}_n' \times (1, 0, 0)\|}, \]
and the skew-symmetric matrix $\mathbf{K}$ is expressed as:
\[
\mathbf{K} = \begin{bmatrix}
0 & -a_z & a_y \\
a_z & 0 & -a_x \\
-a_y & a_x & 0
\end{bmatrix}.
\]

The rotation angle $\theta$ is:
\[ \theta = \arccos\left(\frac{\mathbf{p}_n'}{\|\mathbf{p}_n'\|} \cdot (1, 0, 0)\right). \]

3) Scaling: Scale the set of points so that the x-coordinate of the endpoint becomes 1. The scaling factor is calculated as:
\[
s = \frac{2}{\mathbf{p}_n^{\prime\prime}[0]}.
\]

After scaling, the set of points on the curve is:
\[
\mathbf{P}'' = s \cdot (\mathbf{R} \cdot \mathbf{P}') - (1, 0, 0).
\]

Through these steps, the curve is normalized to a standard form with the starting point at $(-1, 0, 0)$ and the endpoint at $(1, 0, 0)$.

\begin{figure}[t]
    \centering
    \includegraphics[width=0.99\linewidth]{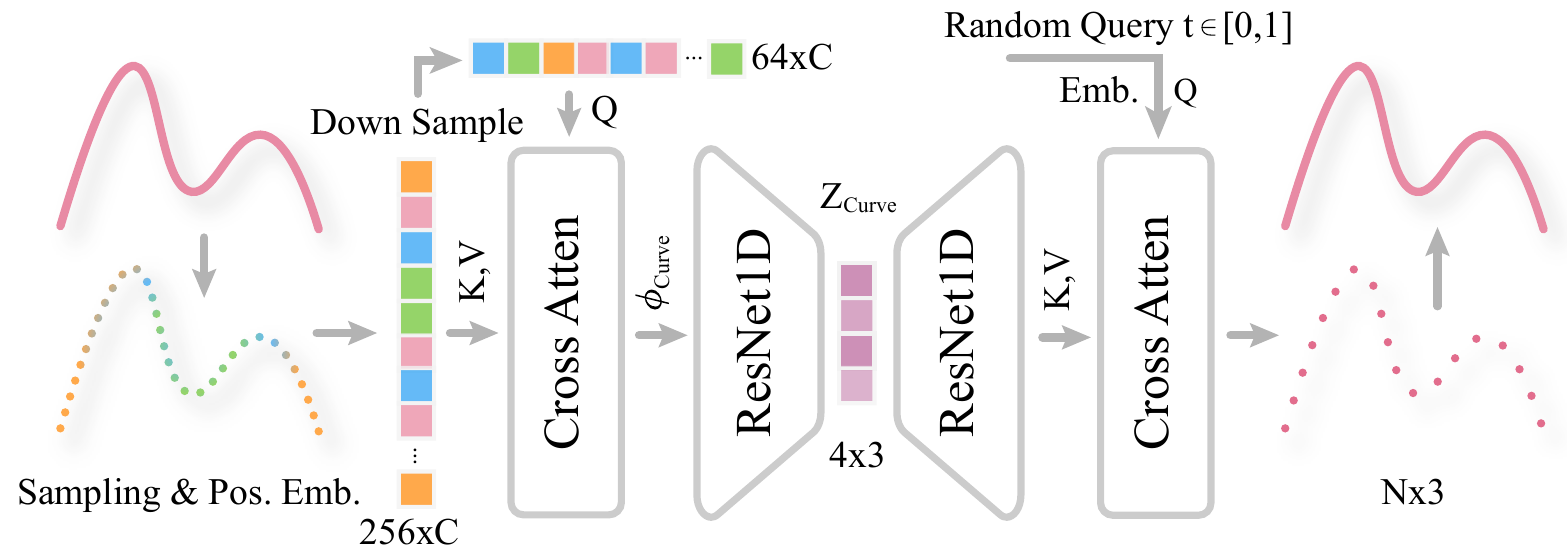}
    \caption{
        CurveVAE training pipeline: Input curves are sampled and positionally encoded into point embedding sequences, refined into $\phi_\text{Curve}$ via cross-attention, and encoded into latent space $Z_\text{Curve}$. These are upsampled and decoded with a random query $t$ to reconstruct the curves.
    }
    \label{fig:curve_vae}
\end{figure}

\paragraph{Training.} We utilize CurveVAE to encode the normalized curves and obtain their neural paramatric representations.
The training process of CurveVAE is illustrated in Fig.~\ref{fig:curve_vae}: First, the input curves are converted into point embedding sequences ($256 \times C$) through sampling and positional encoding. These point sequences are then processed by a cross attention module, which transforms them into compact representations $\phi_\text{Curve} \in \mathbb{R}^{64\times C}$. Subsequently, $\phi_\text{Curve}$ are passed through a ResNet1D~\cite{resnet} encoder to extract both local features and global information, resulting in a more abstract representation of the curves $Z_\text{Curve} \in \mathbb{R}^{4\times3}$. 
Next, $Z_\text{Curve}$ are upsampled using a ResNet1D~\cite{resnet} module.
Then, a random query $(t \in [0,1])$ is introduced and embedded, followed by another cross attention module that decodes the latent representations into the reconstructed curves.
This process compresses the complexity of the curves, while ensuring the continuity and smoothness of the reconstructed curves. 
As a result, the model learns to represent different curves as neural parametric curves within the latent space.

\subsection{Adjacency List in Wireframe VAE}

\begin{figure}[t]
    \centering
    \includegraphics[width=0.99\linewidth]{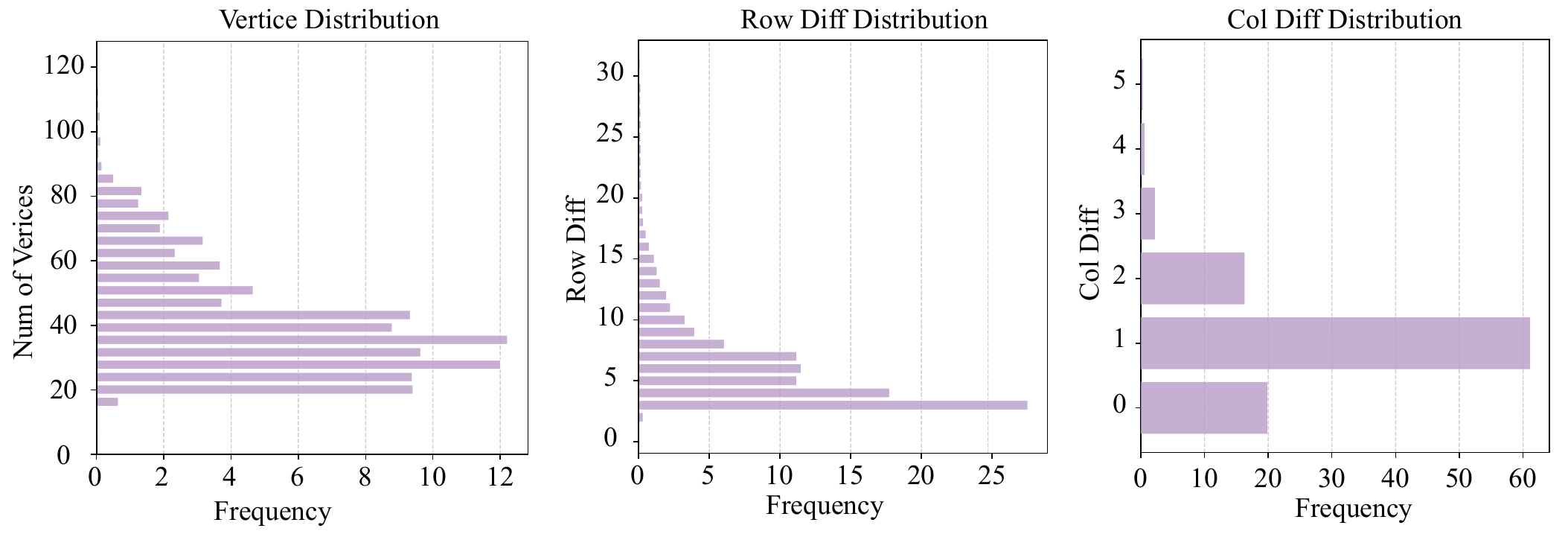}
    \caption{
        Distribution of wireframe vertex counts, row differences (Row Diff) and column differences (Col Diff) in the dataset.
    }
    \label{fig:diff_distribution}
\end{figure}
\begin{figure}[t]
    \centering
    \includegraphics[width=0.96\linewidth]{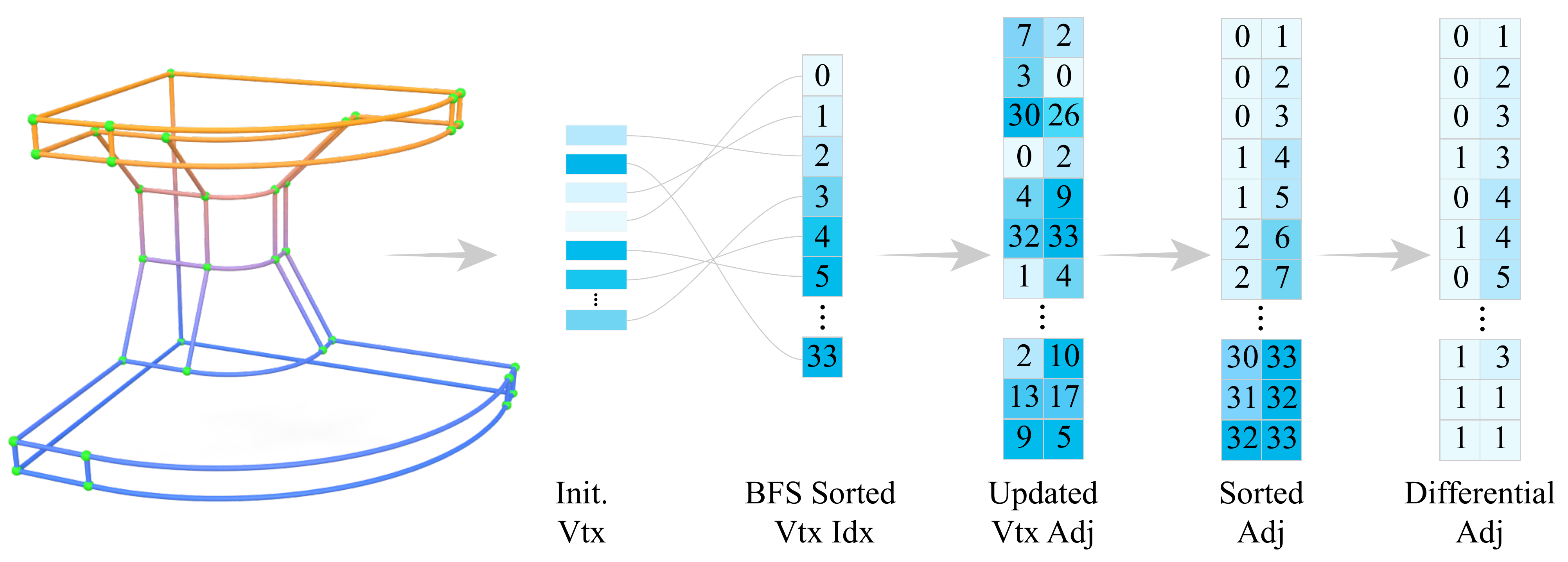}
    \caption{
        The construction process of the differential adjacency list $\Delta \text{Adj}_V$.
        This transformation compresses the adjacency list, streamlining element distribution and enhancing network learning efficiency.
    }
    \label{fig:diff_adjs}
\end{figure}

\begin{figure}[t]
    \centering
    \includegraphics[width=0.99\linewidth]{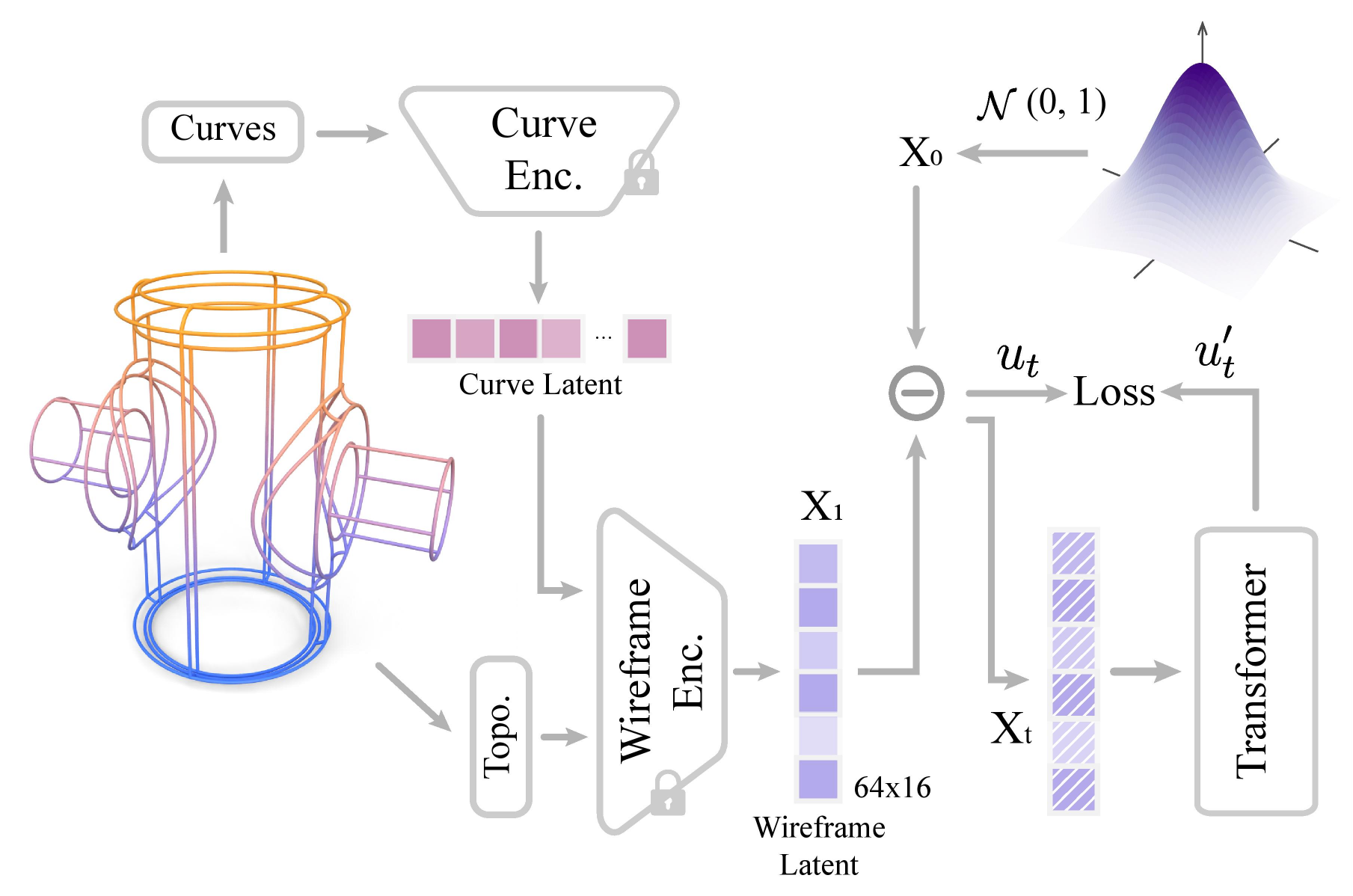}
    \caption{
        The wireframe is mapped to the latent space using Curve and Wireframe encoders. Latent flow matching model then aligns the wireframe latent with a Gaussian noise distribution to generate samples that match the target distribution.
    }
    \label{fig:lfm}
\end{figure}

The computation process of the differential adjacency list $\Delta \text{Adj}_V$ is illustrated in Fig.~\ref{fig:diff_adjs}. 
First, following the 3DWire~\cite{3dwire}, we sort the vertices of the curve wireframe in the z-y-x order to establish an initial organization. 
Then, we apply a breadth-first search (BFS) to perform a further sorting of the vertices, which ensures that topologically closer vertices have proximate index values. 
Next, we update the vertex indices in the curve endpoint adjacency list, and sort the adjacency list in ascending order. 
Finally, we compute the differential adjacency list $\Delta \text{Adj}_V$, by performing a differential calculation on the sorted adjacency list.

We observed that the differential adjacency list $\Delta \text{Adj}_V$ effectively compresses the storage space of the adjacency list while improving computational efficiency.
As shown in Fig.~\ref{fig:diff_distribution}, We analyzed the distribution of the number of vertices in the wireframes of the dataset, as well as the distribution of the first column (Col Diff) and second column (Row Diff) elements in the differential adjacency list $\Delta \text{Adj}_V$. 
Additionally, to address the issue of class imbalance in the element distribution of the differential adjacency list, we employed Focal Loss ~\cite{focal-loss} during training.
In our training process, we set the number of categories for Col Diff to $6$ and the number of categories for Row Diff to $24$.

\subsection{Latent Flow Matching}

The training process of latent flow matching is shown in Fig. ~\ref{fig:lfm}. 
First, all curves in the wireframe are encoded using the Curve Encoder, extracting the curve latent representation $Z_\text{Curve} \in \mathbb{R}^{4\times3}$ of each curve.
These curve latent representations are flattened, combined with topological information, and then fed into the Wireframe Encoder to extract the wireframe latent representation $Z_W \in \mathbb{R}^{64\times16}$.
Next, $Z_W$ is mapped to an initial latent variable $X_0$, which is sampled from a standard normal distribution $\mathcal{N}(0,1)$. 
Using a Transformer module and progressively adjusted time steps $t$, the model learns the velocity $u_t$ corresponding to the target latent variable $X_t$, and calculates a matching loss between the predicted and target velocities. 
This loss guides the model to learn a latent flow in the latent space, effectively capturing the complex mapping between the Gaussian distribution and the wireframe latent representation.
Through this process, the model achieves efficient and smooth generation of wireframe latent representations.

\subsection{Implementation Details}
The network architecture consists of CurveVAE, WireframeVAE, a Latent Flow Matching Model~\cite{flowmatching}, and various encoders for different conditional inputs.
\begin{itemize}
    \item \textbf{Curve VAE.} Both the encoder and decoder utilize $2$ cross-attention layers, with each layer having a hidden channel size of $C=64$. The U-Net backbone for CurveVAE includes $4$ DownBlocks and $4$ UpBlocks, each comprising $2$ layers, with the channels set to $[64, 128, 256, 512]$.
    \item \textbf{Wireframe VAE.} The encoder consists of $4$ cross-attention layers, while the decoder features $12$ self-attention layers followed by $2$ cross-attention layers. All attention layers in WireframeVAE have a hidden channel of $C_w=768$ and $12$ heads, with the wireframe latent space representation is defined as $64\times16$.
    For feature embeddings, we set $Emb_c = 128$, $Emb_v = 2\times 3 \times 128 = 768$, Col Diff embedding to $16$, Row Diff embedding to $32$, resulting in a total adjacency embedding $Emb_a = 48$.
    \item \textbf{Latent Flow Matching Model.} It adopts DiT~\cite{dit} as its backbone, incorporating $12$ attention layers with a hidden channel size of $768$ and $12$ heads.
    \item \textbf{Encoders for Conditional Inputs.} PointNet++ ~\cite{pointnet++} is the encoder for point cloud conditions, while DINOv2 ~\cite{dinov2} for single-view image and sketch conditions. The output features from these encoders are injected into the flow matching model using the AdaLN-Zero~\cite{dit}.
\end{itemize}
All models were trained on a single server equipped with eight NVIDIA 4090 GPUs. 
Additionally, in latent space interpolation, linear interpolation may cause magnitude shrinkage due to high-dimensional vectors' near-orthogonality. We use spherical linear interpolation (slerp) to mitigate this, ensuring smooth transitions.

\section{Experiment Details}
\subsection{Data Processing}
\paragraph{Various conditions.}
We introduced multiple input modalities to comprehensively evaluate our model across various conditional generation tasks.
For single-view and sketch conditions, models were rendered into $224\times224$ pixel images using the OpenCascade Viewer, capturing $24$ random viewpoints for each model. For point cloud-based conditions, we randomly sampled $1024$ points from the surface of each model to generate sparse point clouds. Additionally, to simulate realistic scanned partial point clouds, the Hidden Point Removal algorithm ~\cite{hpr07} was applied.

\paragraph{Data augmentations.}
To enhance the model’s generalization ability, we apply several data augmentation techniques, including rotation, mirroring, random translation, and scaling. For rotation and mirroring, we rotate samples in $90^\circ$ increments along the $x$, $y$ and $z$ axes and mirror them across the $xy$, $xz$, and $yz$ planes. This process yields a total of $4\times4\times4\times3 = 192$ transformations. After removing duplicates, we retain $48$ unique transformations. 
The scaling adjustments were restricted to the range of \( [0.9, 1.1] \), while translations were applied within the range of \( [-0.1, 0.1] \).

\subsection{Quantitative Evaluation Metrics}

In this section, we elaborate on the quantitative metrics used for evaluating model performance, including Coverage (COV), Minimum Matching Distance (MMD), 1-NN Classifier Accuracy (1-NN), Chamfer Distance (CD), Earth Mover's Distance (EMD), F-score (F1), Corner Chamfer Distance (CCD), and Topology Consistency(Topo.).

\paragraph{Generation.}
For unconditional wireframe generation, we use COV, MMD, and 1-NN.
For convenience, we denote the set of generated wireframes as \( W_g \), and the reference wireframe as \( W_r \).
For each wireframe in the sets, we first convert it into a point cloud by uniformly sampling 2048 points from the curves, and then compute the metrics as defined in ~\cite{pointflow}.

COV measures the proportion of wireframes in the reference set \( W_r \) that can be matched by at least one wireframe in the generated set \( W_g \). It is computed as follows:
\[
\text{COV}(W_g, W_r) = \frac{\left| \left\{ \arg\min_{Y \in W_r} D(X, Y) \,\middle|\, X \in W_g \right\} \right|}{|W_r|}
\]

MMD evaluates the shortest distance between each reference point cloud in \( W_r \) and its nearest neighbor in the generated set \( W_g \). It reflects the geometric similarity between the generated and reference wireframes. The formula is:
\[
\text{MMD}(W_g, W_r) = \frac{1}{|W_r|} \sum_{Y \in W_r} \min_{X \in W_g} D(X, Y)
\]
Where \( D(\cdot, \cdot) \) denotes the distance between two wireframes.

1-NN assesses whether each wireframe's nearest neighbor originates from the generated set \( W_g \) or the reference set \( W_r \), which helps evaluate the distributional consistency between the two sets.First, the nearest neighbor \( N_X \) of a wireframe \( X \) is defined as:
\[
N_X = \arg\min_{Y \in W} D(X, Y)
\]
Where \( W \) is the union of \( W_g \) and \( W_r \), excluding \( X \) itself.
The 1-NN metric is then computed as:
\[
1\text{-}\text{NN}(W_g, W_r) = \frac{\sum_{X \in W_g} \mathbb{I}[N_X \in W_g] + \sum_{Y \in W_r} \mathbb{I}[N_Y \in W_r]}{|W_g| + |W_r|}
\]
Where:
\( \mathbb{I}[\cdot] \) is the indicator function, which equals 1 if the condition is true and 0 otherwise.

For point cloud condition generation, we use CD, EMD, and F-score (F1) metrics to evaluate the quality of wireframe reconstruction from the point cloud.

\paragraph{Ablation studies.}
Apart from Chamfer Distance (CD), Earth Mover's Distance (EMD), and F-score (F1), we additionally introduce Corner Chamfer Distance (CCD) and Topology Consistency(Topo.) to assess the reconstruction quality of wireframes. These additional metrics are designed to evaluate the precision of vertex reconstruction and the consistency of the wireframe's topological structure, respectively.
For Topology Consistency, we compare the topological structure of reconstructed wireframe samples with the ground truth (GT) wireframes to determine whether they are isomorphic. The proportion of isomorphic wireframes in the test set is then calculated as a quantitative indicator of topological consistency. This approach effectively evaluates the model's ability to preserve structural relationships in the wireframes.

\section{More Evaluations and Analysis}

\subsection{Ablation Study}
\begin{table}[t!]
    
    \caption{
    Analysis of normalization on curve reconstruction. With the curve normalization, the model achieves significant improvements across all metrics.
    CD is multiplied by \(10^5\) and EMD is multiplied by \(10^3\). 
    }
    \label{tab:ablation_curve}
    \centering
    \setlength{\tabcolsep}{5pt}
    {
    \begin{tabular}{@{}lcccccc@{}}
        \toprule
        Method & CD(\textdownarrow) & EMD$\downarrow$ & F-score$\uparrow$\\ 
        \midrule
        w/o Normalize      & 5.02 & 3.73 & 0.947  \\
        w/ Normalize       & \textbf{1.91} & \textbf{1.24} & \textbf{0.995} \\
        \bottomrule
        
    \end{tabular}
    }
\end{table}

\begin{table}[t!]
    
    \caption{
        Ablations of our design choices on the ABC dataset. 
        (A) with differential adjacency list $\Delta \text{Adj}_V$.
        (B) without differential adjacency list.
        (C) with neighbor-vtx fusion.
        The Dim column represents the Wireframe VAE Latent Dim. 
        The results indicate a significant decrease in performance when any of them are removed. 
        CD is multiplied by \(10^4\) and EMD is multiplied by \(10^3\).
    }
    \label{tab:ablation_wf}
    \centering
    \setlength{\tabcolsep}{5pt}
    \resizebox{\linewidth}{!}
    {
    \begin{tabular}{@{}c|c|ccccc@{}}
        \toprule

        Setting & Dim & CD (\textdownarrow) & EMD (\textdownarrow) & F1 (\textuparrow) & \makecell{CCD (\textdownarrow)} & \makecell{Topo.}(\% \textuparrow) \\ 
        \midrule
        \multirow{2}{*}{A} & 4  & 2.45 & 7.26 & 0.607 & 3.85 & 99.29 \\
                                               & 8  & 1.43 & 7.03 & 0.817 & 2.35 & 99.72 \\
        \midrule
        B & \multirow{2}{*}{16} & 1.15 & 6.67 & 0.860 & 1.72 & -     \\
        C &  & 0.98 & 6.59 & 0.895 & 1.51 & 77.84 \\
        \midrule
        A & 16 & \textbf{0.81} & \textbf{6.48} & \textbf{0.926} & \textbf{1.28} & \textbf{99.91} \\
        \bottomrule
        
    \end{tabular}
    }
\end{table}

We conduct ablation study to analyze the contributions of different components.
Specifically, we focused on the following aspects: 1) Curve Normalization; 2) Difference Adjacency $\Delta \text{Adj}_V$; and 3) Dimension of the wireframe Latent Space (Latent Dim).
We use the CD, EMD, F-score to evaluate the reconstruction accuracy of the curves and wireframe, Corner CD (CCD) to evaluate the reconstruction accuracy of the wireframe corner, and the topology consistency rate (Topo.) to evaluate the topological consistency of the wireframe.

\begin{figure}[t]
    \centering
    \includegraphics[width=0.99\linewidth]{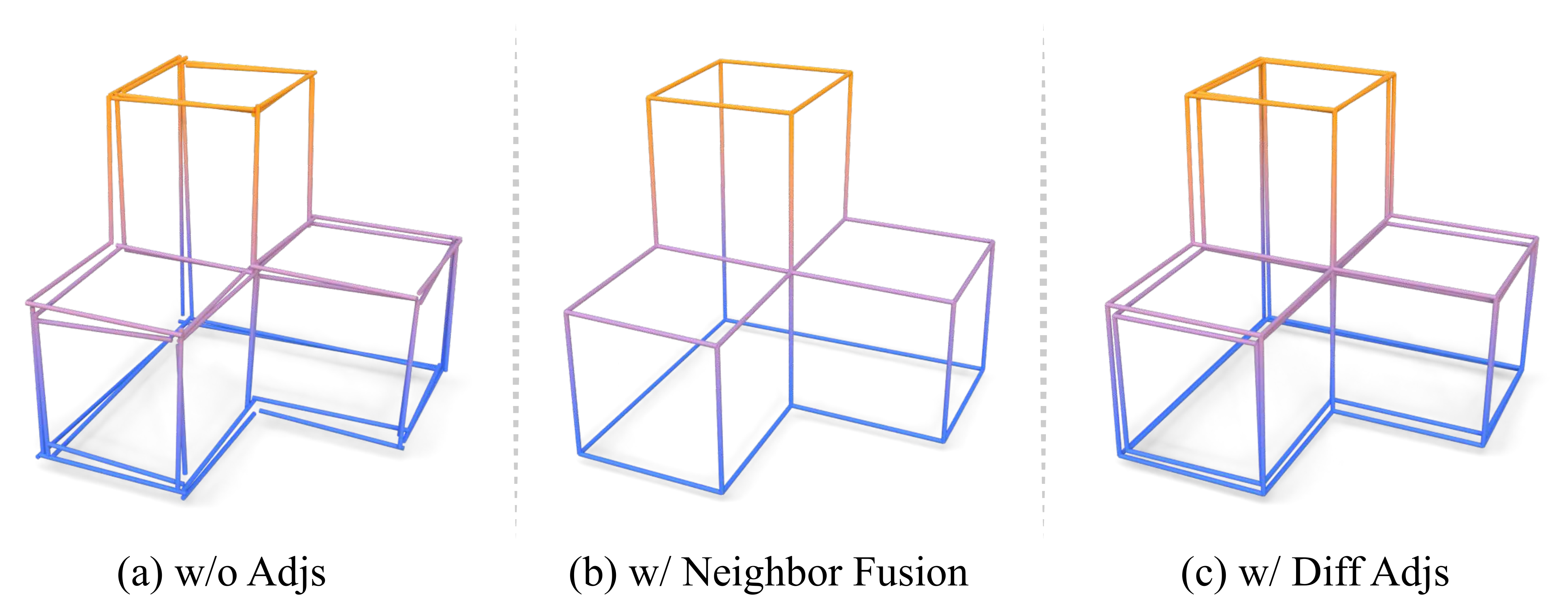}
    \caption{
        Ablation study on the wireframe reconstructed by (a) removing the adjacency constraint; (b) replacing the adjacency constraint with the neighbor vertex fusion; and (c) using our differential adjacency constraint.
    }
    \label{fig:ablation_diff}
\end{figure}

The quantitative results, as shown in Table~\ref{tab:ablation_curve}, demonstrate that this normalization process significantly improves reconstruction accuracy, which we attribute to the relatively constrained and stable nature of the curve space.
Additionally, we evaluated the effect of Difference Adjacency $\Delta \text{Adj}_V$ and wireframe latent dimension, as presented in Table \ref{tab:ablation_wf}. Removing Difference Adjacency $\Delta \text{Adj}_V$ led to a notable decline in all evaluated metrics, especially in wireframe reconstruction and corner reconstruction, demonstrating its importance in modeling the local geometric relationships between vertices. 
Alternatively, Neighbor-Vertex Fusion (Neighbor-Vtx Fusion) still fell short of the default (Differential Adjacency) in terms of reconstruction precision.
As shown in Fig.~\ref{fig:ablation_diff}, we visualize the impact of Difference Adjacency $\Delta \text{Adj}_V$ on reconstruction quality.

\begin{figure}[t]
    \centering
    \includegraphics[width=0.99\linewidth]{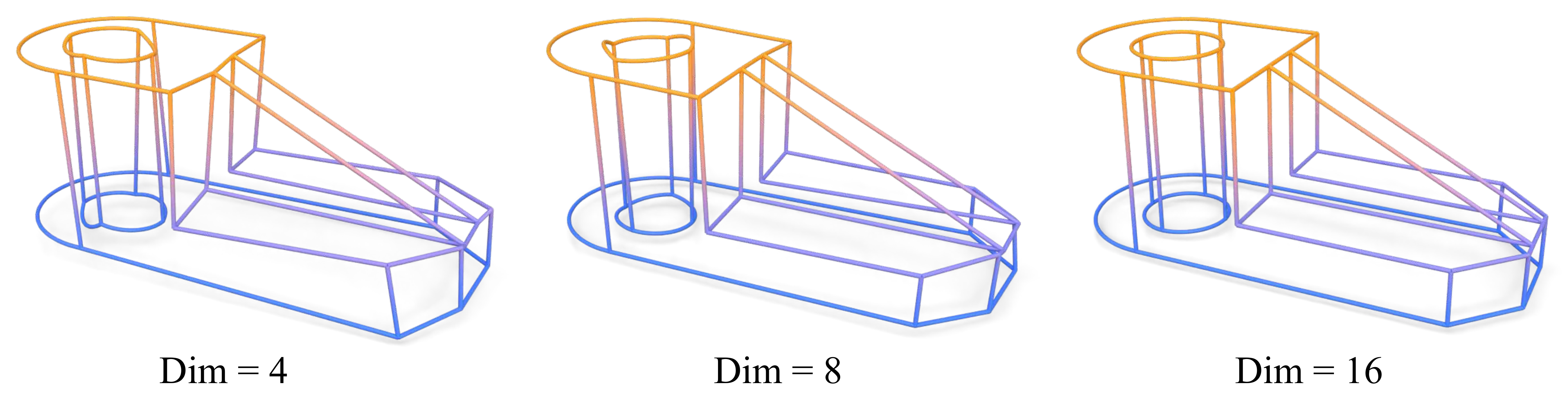}
    \caption{
        Visualization of the impact of different latent space dimensions (Latent Dim) on reconstruction results. 
    }
    \label{fig:ablation_latent_dim}
\end{figure}

For Latent Dim, we compared the default dimension (16) with 4 and 8. The results show that as Latent Dim increases, the reconstruction error decreases significantly, demonstrating our method’s scalability with respect to network hyperparameters.
Furthermore, Latent Dim had a relatively small impact on topology consistency but significantly influenced Corner CD, emphasizing its role in modeling the spatial positions of wireframe vertices. We also include a visualization (Fig.~\ref{fig:ablation_latent_dim}) to show the impact of different Latent Dim settings on reconstruction quality.


\begin{table}[t!]
    \caption{
        Quantitative evaluation of unconditional generation on the DeepCAD dataset~\cite{deepcad}. Our method outperforms others on COV, MMD, and 1-NN.
        Note that CD values are multiplied by \(10^2\).
    }
    \label{tab:quantitative_deepcad}
    \centering
    \setlength{\tabcolsep}{5pt}
    {
    \begin{tabular}{@{}lcccccc@{}}
        \toprule
        Method & COV (\%, \textuparrow) & MMD (\textdownarrow) & 1-NN (\%) \\ 
        \midrule
        DeepCAD \cite{deepcad}      & 39.02 & 2.94 & 70.47  \\
        BrepGen \cite{brepgen}      & 41.33 & 2.61 & 67.25  \\
        Ours  & \textbf{46.19} & \textbf{2.43} & \textbf{61.53} \\
        \bottomrule
        
    \end{tabular}
    }
\end{table}

\begin{figure}[t]
    \centering
    \includegraphics[width=0.99\linewidth]{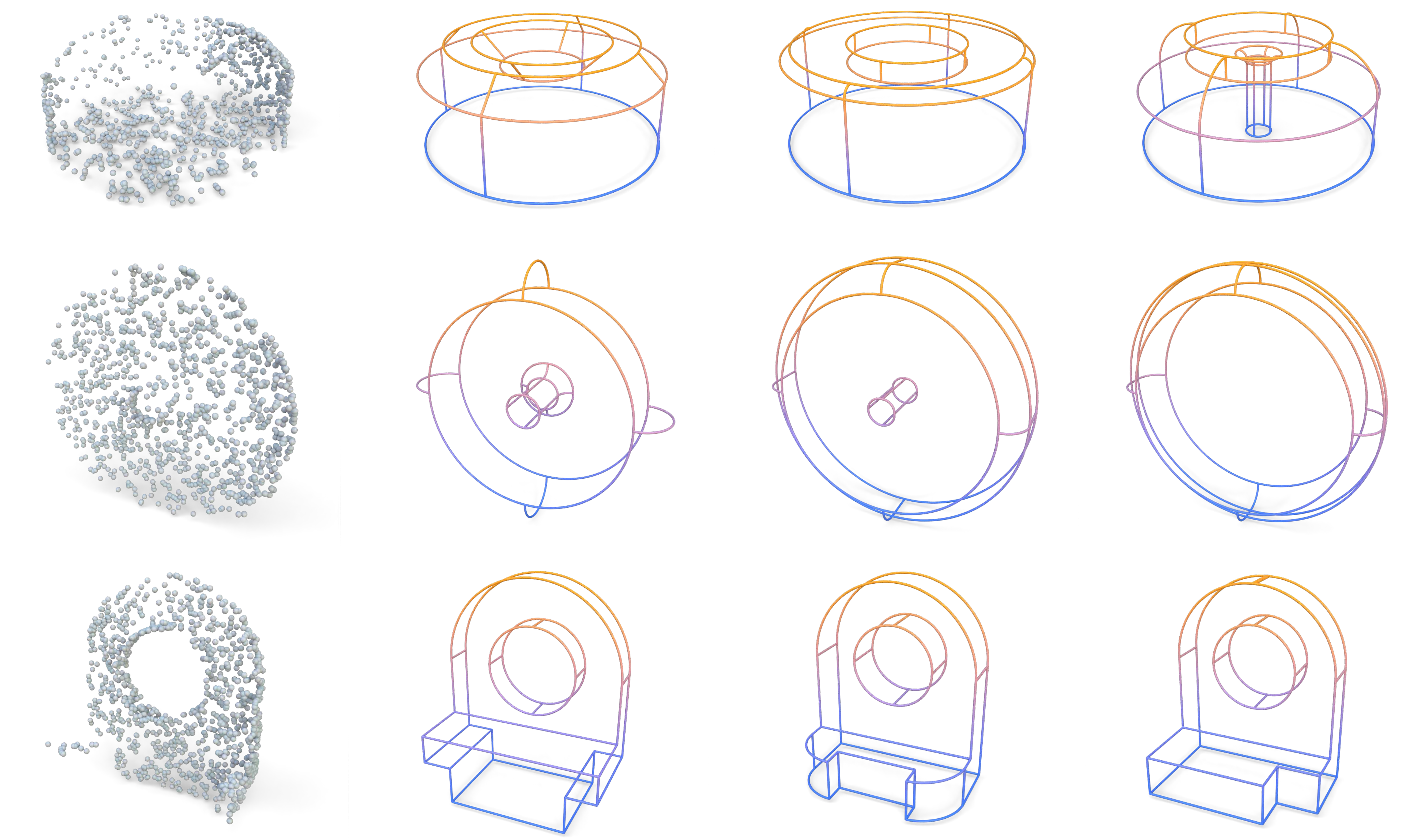}
    \caption{
        From a partial point cloud (left), our method generates multiple wireframes with similar shape but diverse topologies.
    }
    \label{fig:cond_partial_diversity}
\end{figure}
\begin{figure}[t]
    \centering
    \includegraphics[width=0.99\linewidth]{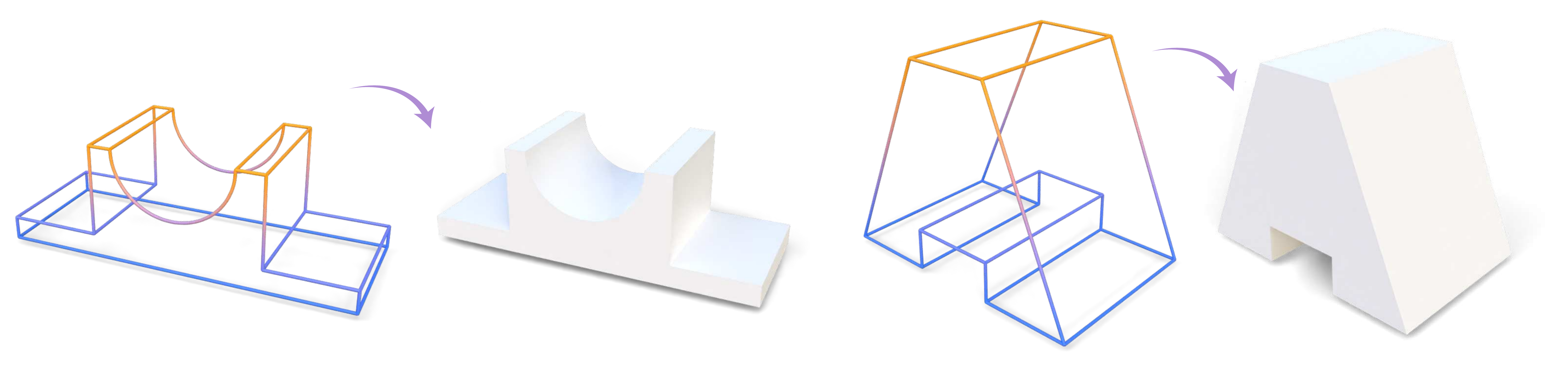}
    \caption{
        The resulting wireframe can be converted into a mesh model.
    }
    \label{fig:wf2mesh}
\end{figure}

\begin{figure}[t]
    \centering
    \includegraphics[width=0.8\linewidth]{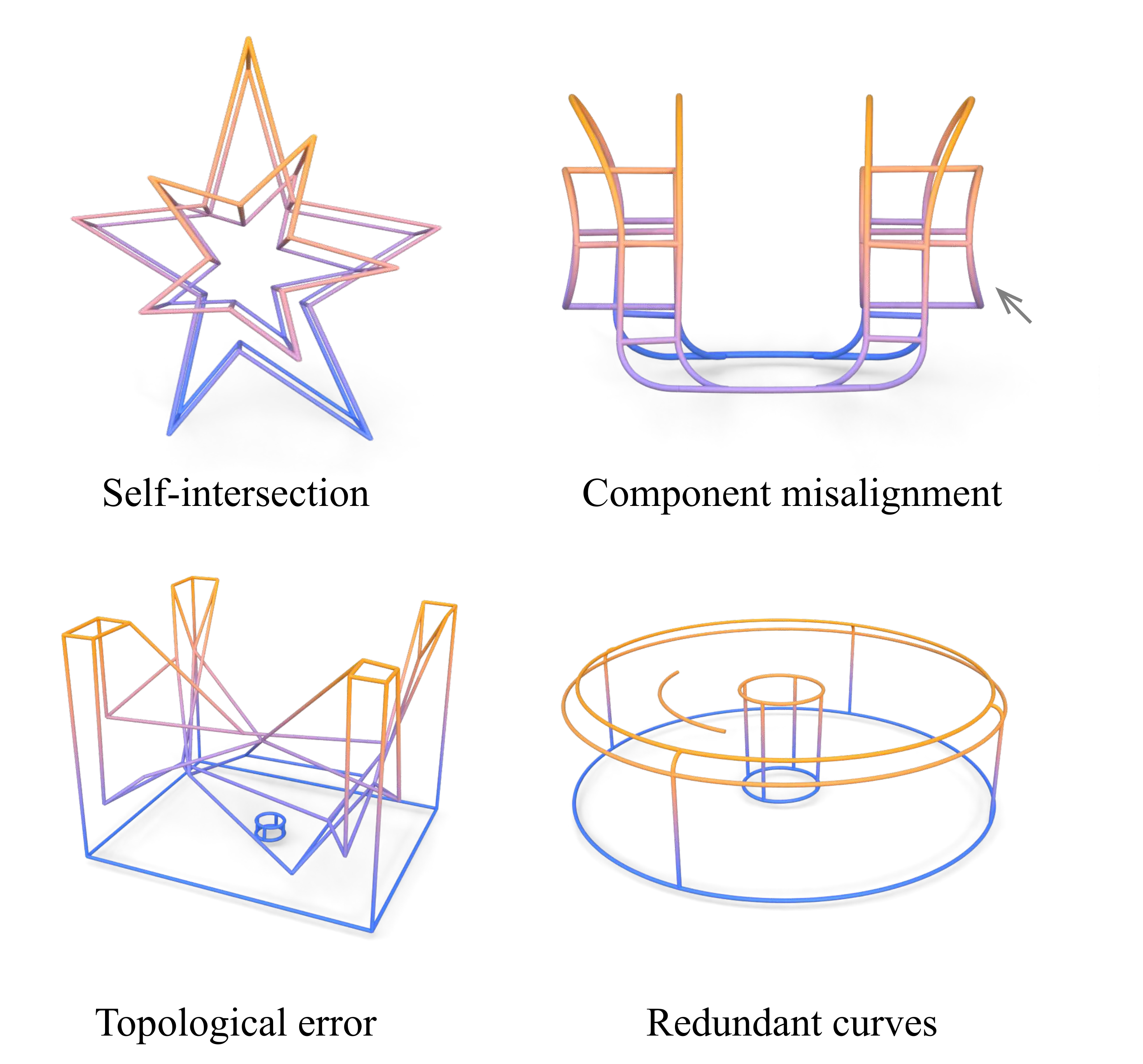}
    \caption{
        Four common failure cases categories for our method.
    }
    \label{fig:failure_case}
\end{figure}

\subsection{Topology Distribution in Latent Space}
To explore the distribution of samples with the same topological structure in latent space, we grouped the wireframes by their topologies and analyzed their clustering.
Given the large size of our dataset, we focused on a subset for consistency. 
Specifically, we selected samples with the most common number of curves ($48$), yielding $6,719$ samples. Within this subset, we identified $574$ distinct topologies.
We evaluated latent space clustering using the silhouette score, a metric that quantifies clustering quality by comparing each sample’s intra-cluster distance (cohesion) to its nearest-cluster distance (separation).
This metric indicates the degree of separation between clusters. 
The silhouette score ranges from $-1$ to $1$, where higher values indicate better clustering. A score near $1$ suggests well-separated clusters, around $0$ indicates overlapping boundaries, and negative values reflect poor clustering.
We observed a silhouette score of $0.1135$, which shows that samples sharing the same topology tend to cluster together.
Notably, the boundaries between different topologies are not sharply defined, suggesting potential overlaps that indicate shared geometric features or continuous transitions between categories.
This result reveals a key property of wireframe representations: shape is affected not only by topology but also by geometric details, such as vertex positions and curve geometry.

\subsection{Visual Comparisons with Other Methods using Our Results for Retrieval}

\begin{figure*}[b]
    \centering
    \includegraphics[width=0.93\linewidth]{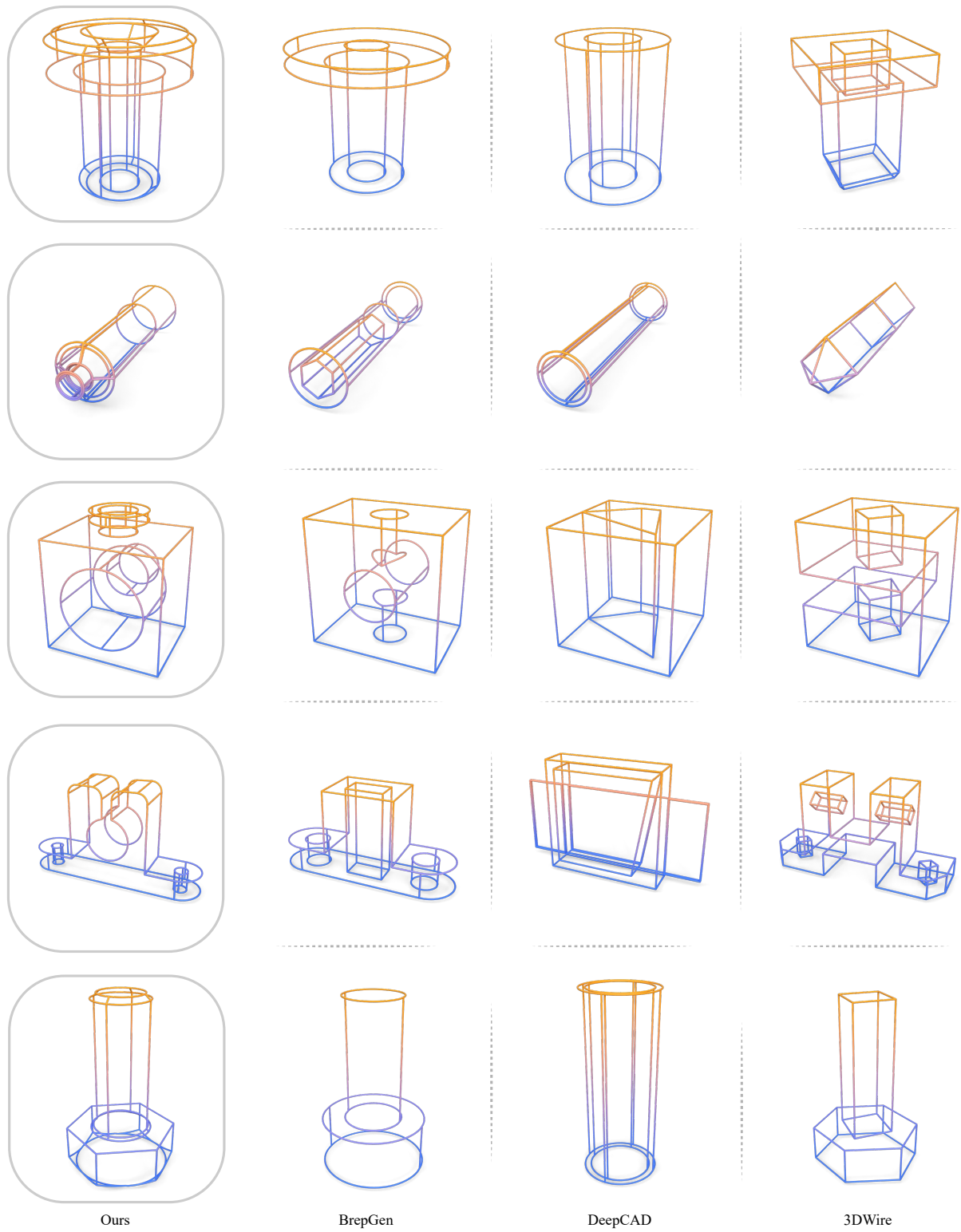}
    \caption{
        Unconditional wireframe generation of different methods.
    }
    \label{fig:ours2others}
\end{figure*}

Fig. \ref{fig:ours2others} shows some visual comparisons. For each random wireframe generated by our method, we find a similar wireframe generated by each of the state-of-the-art methods~\cite{brepgen, deepcad, 3dwire} to make the visual comparison easier.
From all these results shown in Fig. \ref{fig:ours2others}, we can see that the 3D wireframes generated by our method clearly exhibit more complex structures than those generated by the other methods.

\subsection{Visual Comparisons on Our Results Retrieved by BrepGen's Results}

\begin{figure*}[b]
    \centering
    \includegraphics[width=0.95\linewidth]{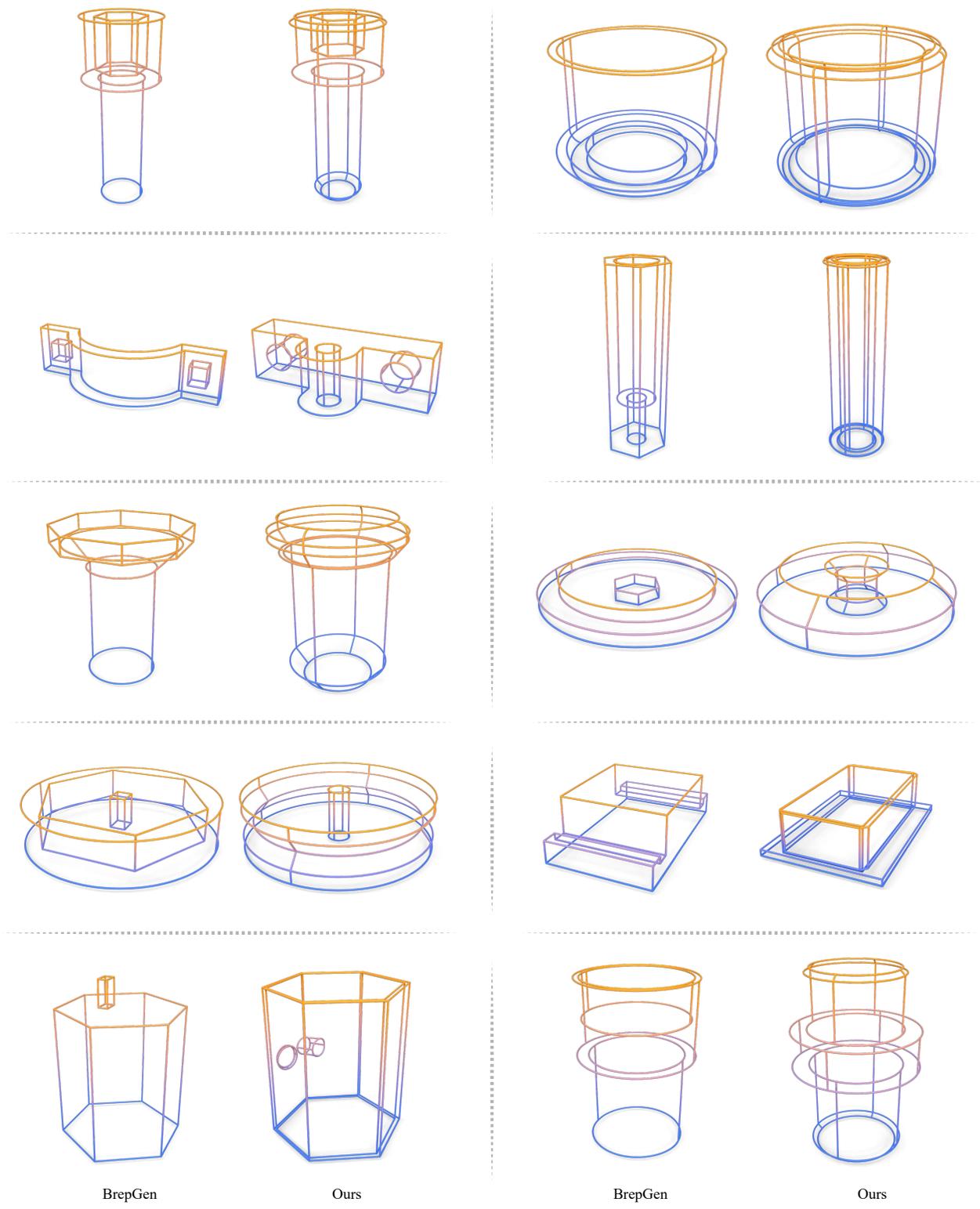}
    \caption{
        Unconditional wireframe generation of different methods.
    }
    \label{fig:brepgen2ours}
\end{figure*}

As different methods potentially have different statistical models in the wireframe generation distribution, in addition to visual comparisons presented earlier in Fig. \ref{fig:ours2others}, we take wireframes generated by BrepGen~\cite{brepgen} and find the most similar wireframe generated by our method for visual comparison. 
From the retrieved results shown in Fig. \ref{fig:brepgen2ours}, we can see that our generated wireframes are richer in details compared to those generated by BrepGen, especially at the chamfer positions of the objects.

\subsection{More Examples for Shape Novelty Analysis}

We present additional examples for shape novelty analysis beyond Fig. 6 in the main paper.
For each wireframe we generated, we retrieved the top-4 from the training set based on Chamfer Distance (CD).
The result shows that they share similar structures, demonstrating that our method can generate realistic wireframes similar to those in the training set. Moreover, our shapes exhibit distinct differences in various local structures.

\subsection{More Results}

We further evaluated our method on the DeepCAD dataset, which is provided by BrepGen, as show in Table \ref{tab:quantitative_deepcad}. Our method achieves superior performance in terms of COV, MMD, and 1-NN metrics, outperforming both DeepCAD\cite{deepcad} and BrepGen\cite{brepgen} baselines.
We provide additional results of interpolation in the latent space, as shown in Fig.~\ref{fig:latent_interpolation_supp}, illustrating how our method smoothly transitions between 3D wireframes of different structures within the latent space.
In Fig.~\ref{fig:cond_sparse_supp}, we showcase more results generated under sparse point cloud condition. This highlights the ability of our method to successfully create well-structured 3D wireframes, even in cases where the point cloud is sparse, particularly for objects with smoother surfaces.
More results under partial point cloud condition, as shown in Fig.~\ref{fig:cond_partial_supp}, demonstrate that our method can still produce complete and consistent 3D wireframes, even when part of the point cloud data is missing.
As shown in Fig.~\ref{fig:cond_partial_diversity}, our method generates multiple wireframes from a partial point cloud input, exhibiting different topologies while preserving overall shape.
Fig.~\ref{fig:cond_image_sketch_supp} displays further results generated from single-view images and sketches. It emphasizes that our method is capable of generating 3D wireframes that align with the overall spatial structure of the input conditions, with a single perspective input such as an image or sketch. 
Additionally, we present more unconditional generation results, as shown in Fig.~\ref{fig:more_uncond_1} and Fig.~\ref{fig:more_uncond_2}.
Furthermore, for simple wireframes, mesh models can be obtained by converting them into graphs and identifying coplanar minimal cycles, as shown in Fig. ~\ref{fig:wf2mesh}. Additionally, for complex wireframes, curve-based surface models can be constructed using existing methods such as \cite{zhuang2013cycles} and \cite{pan2015curve}.
Note that, due to inherent topological ambiguities in complex wireframes, surface reconstruction is not always feasible for all cases.

\section{Discussion on Limitations}

Although our method can generate diverse and structurally complex 3D curve wireframes, it has limitations in recovering fine-grained geometric details when condition is given.
Simply injecting the global feature into the flow-matching process while ensuring high robustness on imperfect data, hinders the model's ability to capture local details.
This issue can also be observed when the input is a single-view image or a sketch.
In these cases, domain drift is exacerbated because our encoder (DINOv2~\cite{dinov2}), which was not pre-trained on the ABC dataset, provides latent features lacking sufficient local geometric information.

While our continuous latent space enables smooth interpolation, intermediate results may exhibit structural inconsistencies (e.g., sudden topology changes), which could be mitigated through regularization of interpolated latent vectors.
In addition, the entangled nature of the latent space limits semantic editability (such as selectively modifying specific components like chair legs or lamp arms), suggesting the need for integrating disentangled semantic features in future work.

Additionally, our method achieves topological alignment in $81.79\%$ of samples (measured by adjacency list and vertex consistency), the remaining cases still face common challenges in previous methods, such as self-intersection, component misalignment, topological errors, and extraneous curves, as shown in Fig. \ref{fig:failure_case}.
To address these issues, we can introduce regularization, spatial consistency constraints, topological graph constraints, or post-processing techniques to improve the integrity of the generated wireframe.

\begin{figure*}[b]
    \centering
    \includegraphics[width=0.99\linewidth]{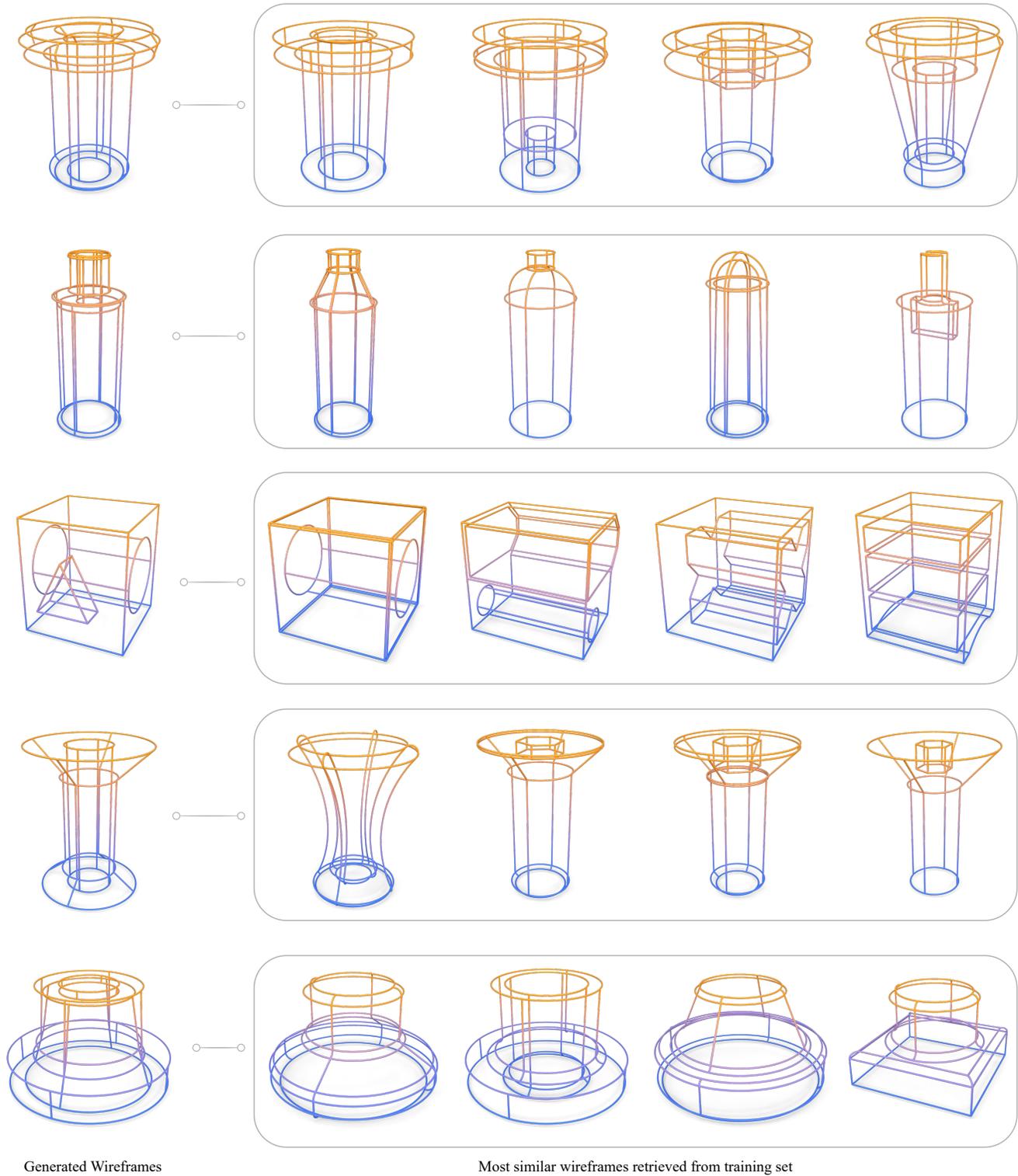}
    \caption{
        Wireframe novelty analysis. For each random wireframe generated by our method, we show the top four most similar wireframes retrieved from the training set by Chamfer Distance (CD).
    }
    \label{fig:novelty_ours}
\end{figure*}

\begin{figure*}
    \centering
    \includegraphics[width=0.96\linewidth]{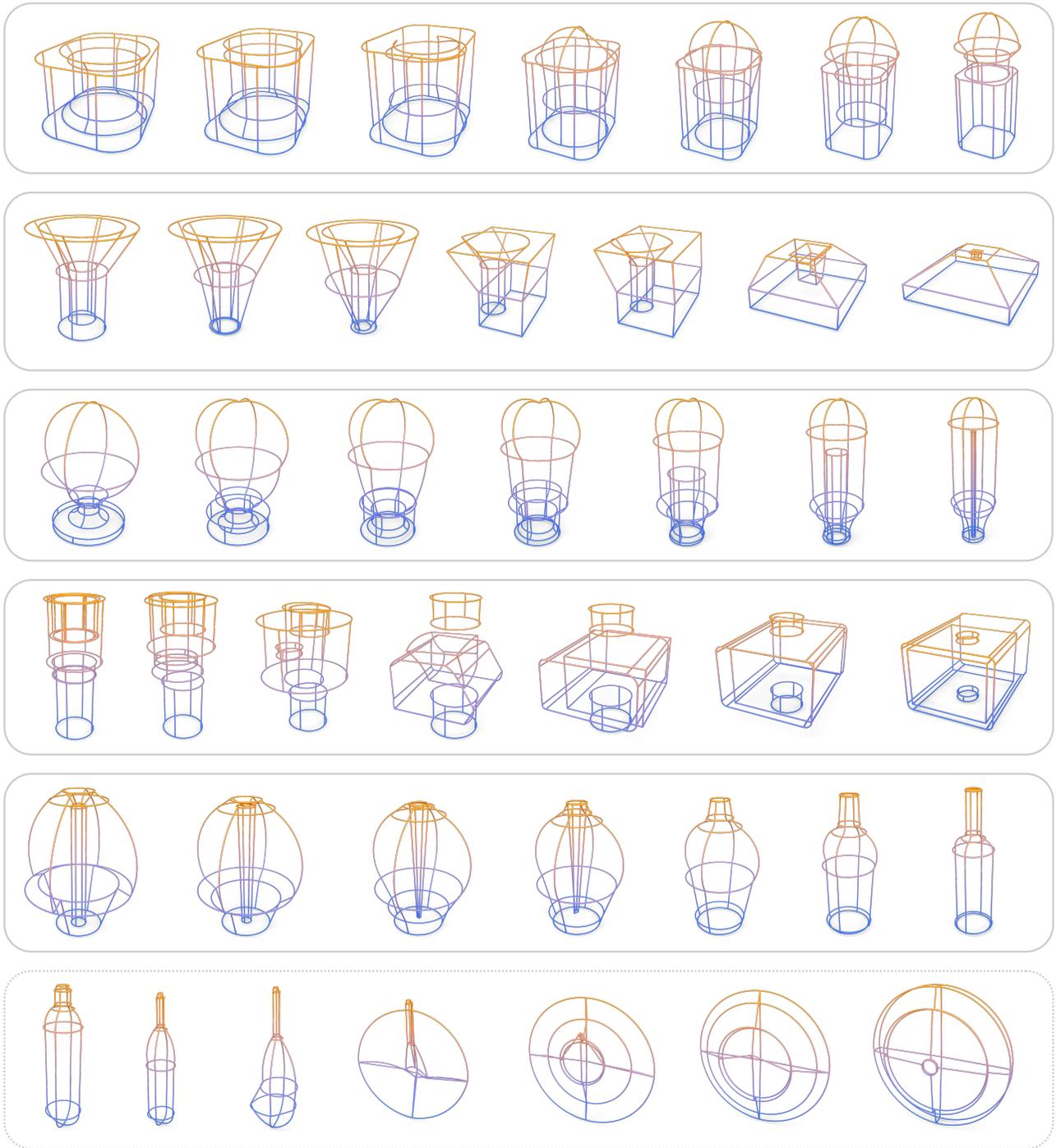}
    \caption{
        Latent Space Interpolation.
    }
    \label{fig:latent_interpolation_supp}
\end{figure*}
\begin{figure*}[t]
    \centering
    \includegraphics[width=0.99\linewidth]{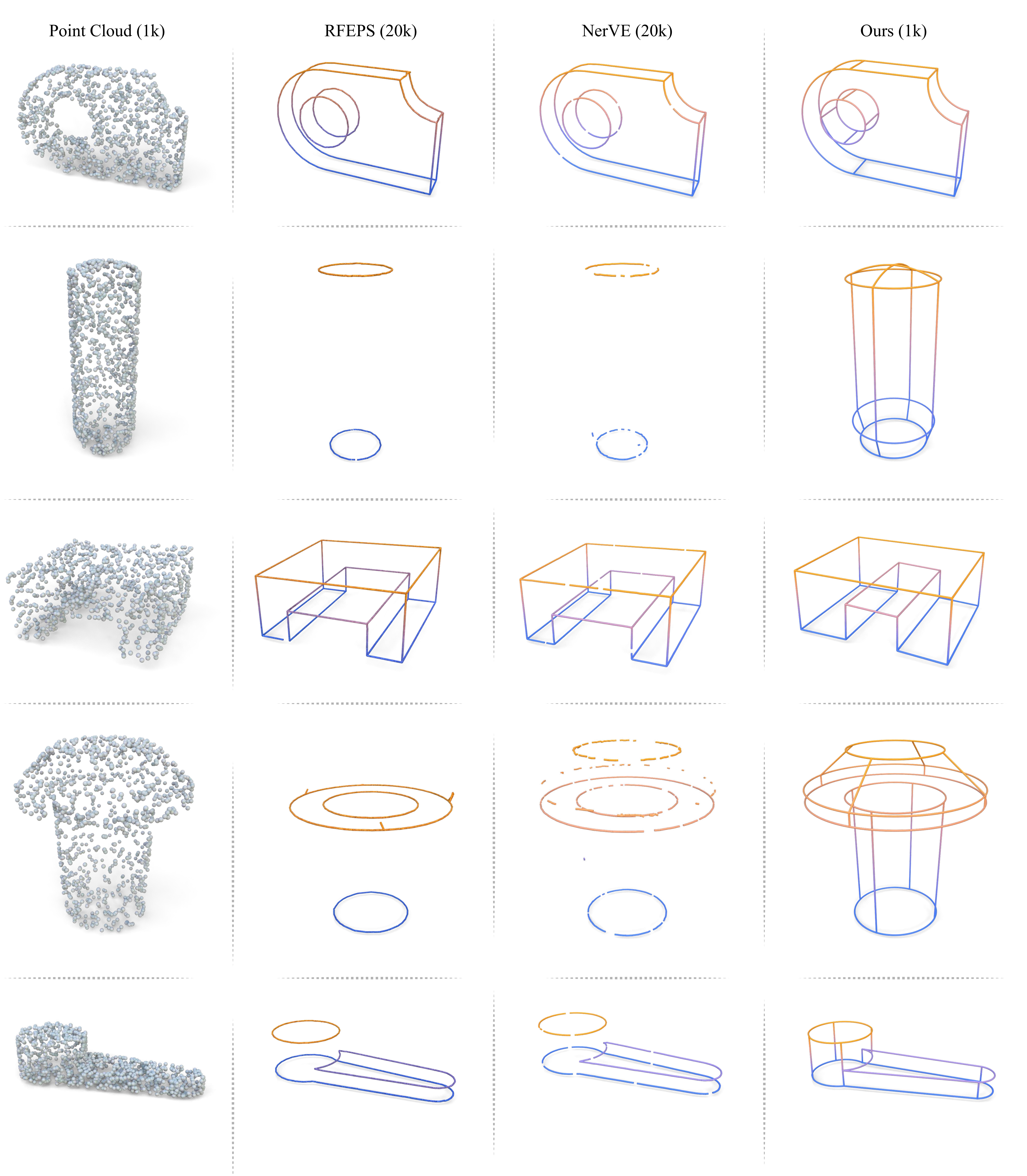}
    \caption{
        More results for sparse point cloud condition generation.
    }
    \label{fig:cond_sparse_supp}
\end{figure*}
\begin{figure*}[t]
    \centering
    \includegraphics[width=0.96\linewidth]{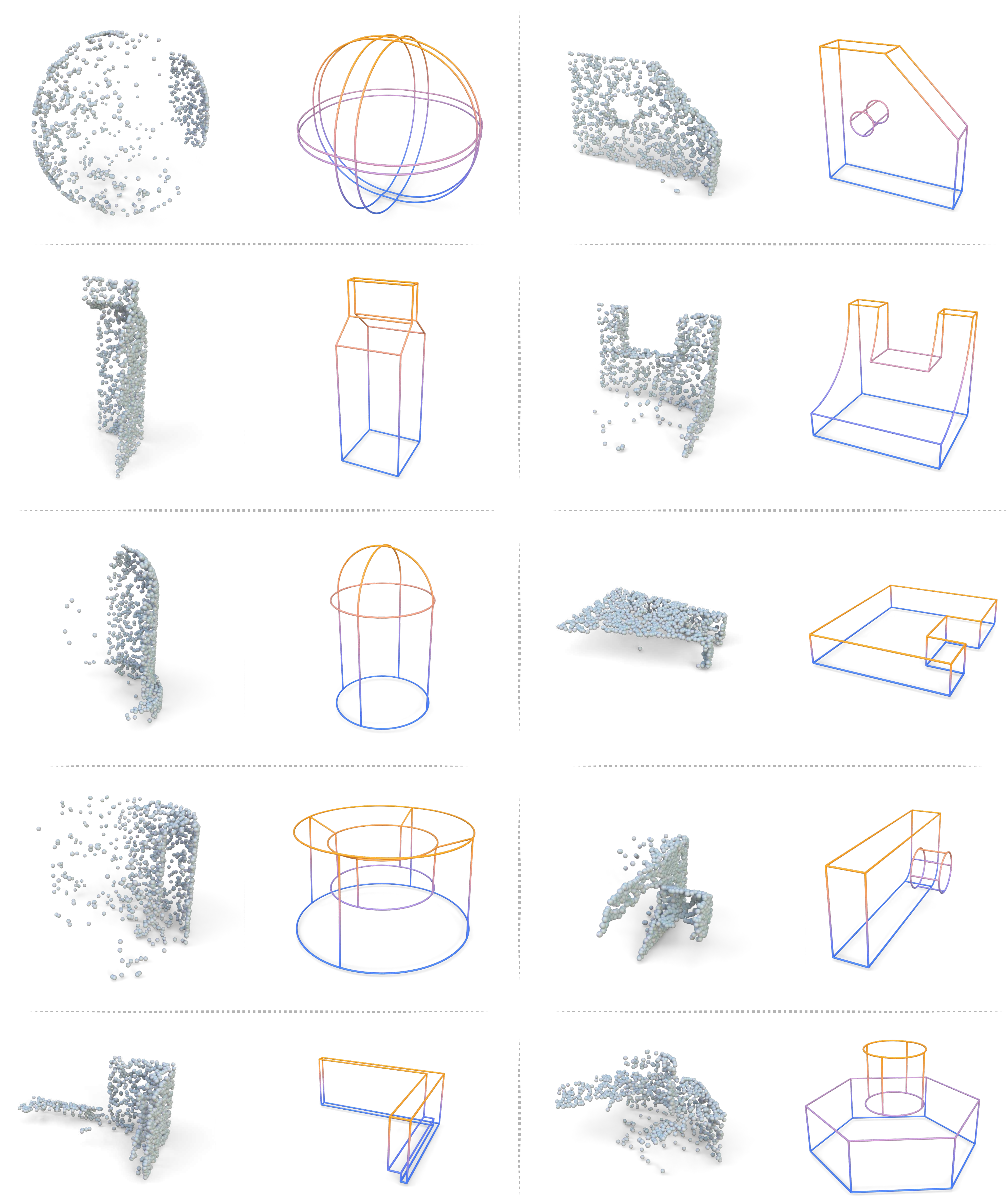}
    \caption{
        More results for partial point cloud condition generation.
    }
    \label{fig:cond_partial_supp}
\end{figure*}
\begin{figure*}[t]
    \centering
    \includegraphics[width=0.96\linewidth]{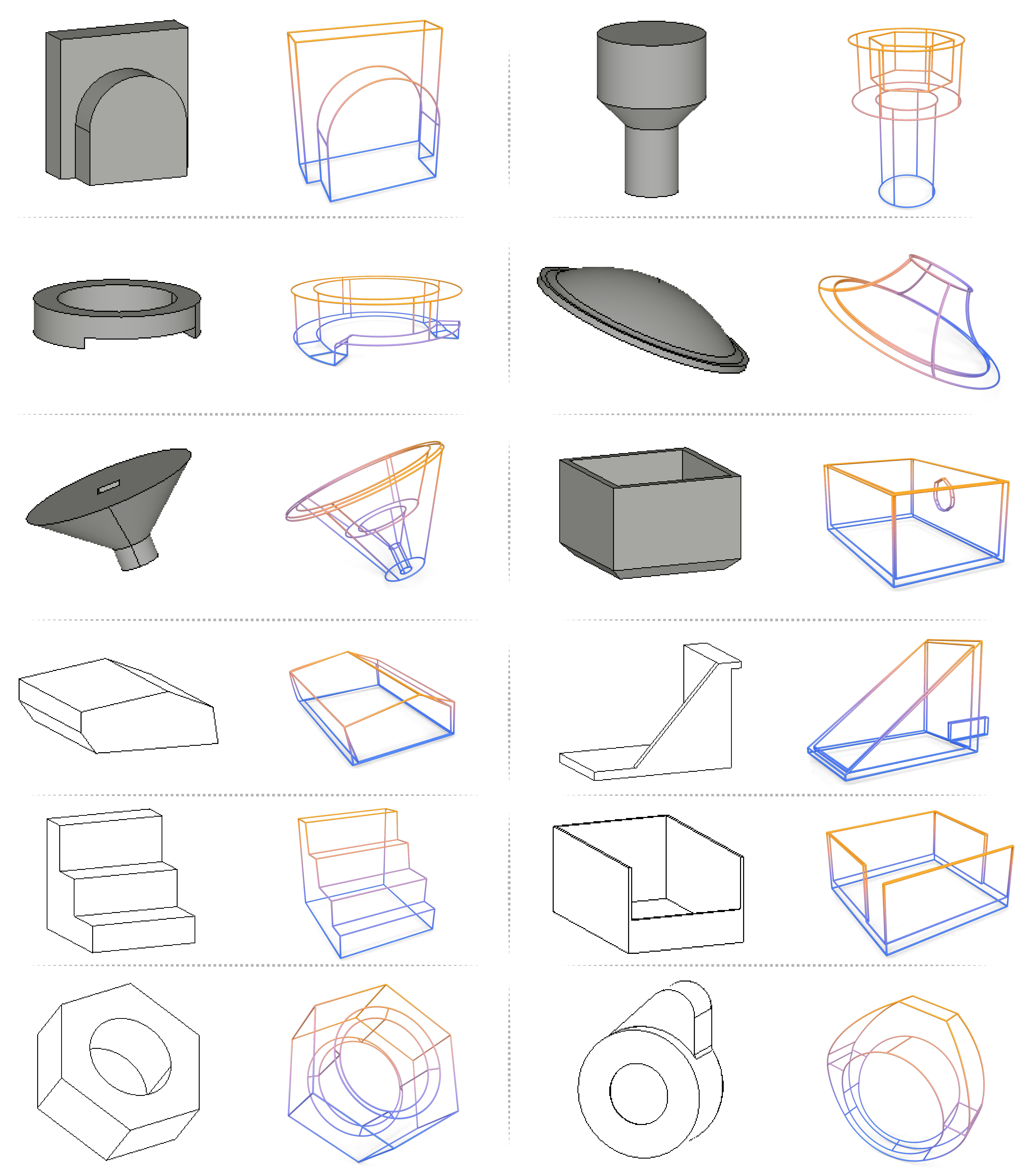}
    \caption{
        More results for single-view image and sketch condition generation.
    }
    \label{fig:cond_image_sketch_supp}
\end{figure*}

\begin{figure*}[hb]
    \centering
    \includegraphics[width=0.99\linewidth]{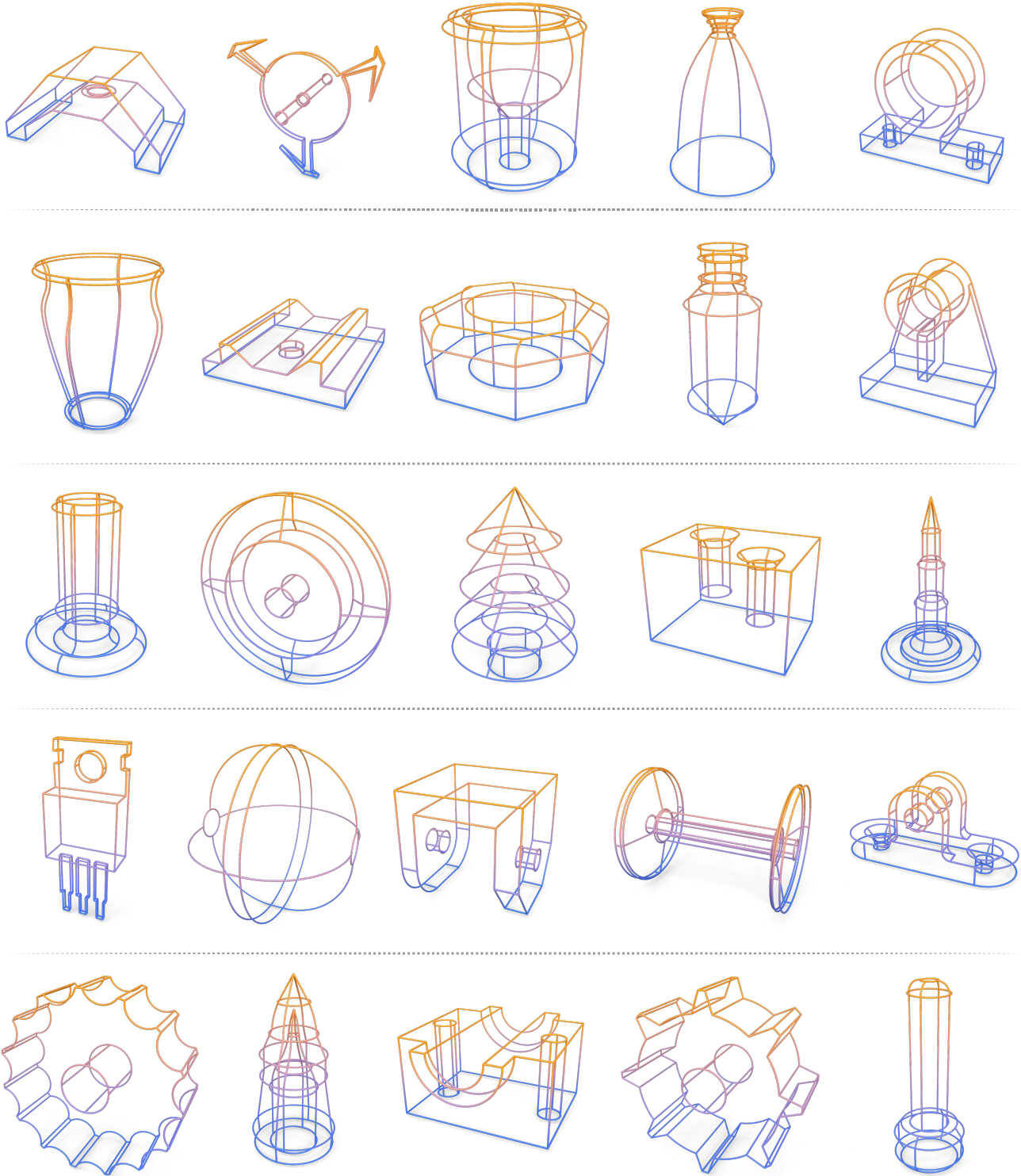}
    \caption{
        More results for unconditional wireframe generation.
    }
    \label{fig:more_uncond_1}
\end{figure*}
\begin{figure*}[hb]
    \centering
    \includegraphics[width=0.99\linewidth]{fig/more_uncond_2_S.pdf}
    \caption{
        More results for unconditional wireframe generation (cont.).
    }
    \label{fig:more_uncond_2}
\end{figure*}

\end{document}